\pdfoutput=1

\documentclass[11pt,review,authoryear]{elsarticle}

\usepackage{amsfonts}
\usepackage{amsmath}
\usepackage{hyperref}
\usepackage{enumitem}
\usepackage[utf8]{inputenc}
\usepackage{bm}
\usepackage{subfigure}
\usepackage{xcolor}
\usepackage{soul}
\usepackage[toc,page]{appendix}
\usepackage{tikz}

\usepackage[margin=1.5cm]{geometry}

\biboptions{authoryear}

\newcommand\copyrighttext{%
  \footnotesize \copyright\ 2021. This manuscript version is made available under the CC-BY-NC-ND 4.0 license \url{http://creativecommons.org/licenses/by-nc-nd/4.0/} This work has been submitted to Physica A for possible publication. Copyright may be transferred without notice, after which this version may no longer be accessible.}
\newcommand\copyrightnotice{%
\begin{tikzpicture}[remember picture,overlay]
\node[anchor=south,yshift=160pt] at (current page.south) {\fbox{\parbox{\dimexpr\textwidth-\fboxsep-\fboxrule\relax}{\copyrighttext}}};
\end{tikzpicture}%
}

\makeatletter
\def\ps@pprintTitle{%
 \let\@oddhead\@empty
 \let\@evenhead\@empty
 \def\@oddfoot{}%
 \let\@evenfoot\@oddfoot}
\makeatother

\begin{document}

\begin{frontmatter}

\title{Minimizing the evacuation time of a crowd from a complex building using rescue guides}

\author[1]{Anton von Schantz}\corref{cor1}
\ead{anton.von.schantz@aalto.fi}
\author[1]{Harri Ehtamo}
\ead{harri.ehtamo@aalto.fi}

\address{Aalto University, Department of Mathematics and Systems Analysis, P.O. Box 11100, FI-00076 Aalto, Finland}

\cortext[cor1]{Corresponding author}

\begin{abstract}
In an emergency situation, the evacuation of a large crowd from a complex building can become slow or even dangerous without a working evacuation plan. The use of rescue guides that lead the crowd out of the building can improve the evacuation efficiency. An important issue is how to choose the number, positions, and exit assignments of these guides to minimize the evacuation time of the crowd. Here, we model the evacuating crowd as a multi-agent system with the social force model and simple interaction rules for guides and their followers. We formulate the problem of minimizing the evacuation time using rescue guides as a stochastic control problem. Then, we solve it with a procedure combining numerical simulation and a genetic algorithm (GA). The GA iteratively searches for the optimal evacuation plan, while numerical simulations evaluate the evacuation time of the plans. We apply the procedure on a test case and on an evacuation of a fictional conference building. The procedure is able to solve the number of guides, their initial positions and exit assignments in a single although complicated optimization. The attained results show that the procedure converges to an optimal evacuation plan, which minimizes the evacuation time and mitigates congestion and the effect of random deviations in agents' motion.
\end{abstract}

\begin{keyword}
Multi-agent systems \sep Numerical simulation \sep Evacuation \sep Stochastic optimization \sep Genetic algorithms
\end{keyword}
\end{frontmatter}

\copyrightnotice

\newpage

\section{Introduction}

Large complex buildings like airport terminals, high-rise buildings and subway stations need crowd management to ensure safety. In an emergency situation, like a fire or a bomb threat, a well-planned operation is needed to ensure a fast and safe evacuation. It is known that people are willing to follow authorities \citep{proulx2002, nishinari}, and that the strategic use of trained safety personnel, or rescue guides, that lead the crowd out of the building, improves evacuation efficiency \citep{leadershipeffect}. However, as the crowd size and complexity of the building increases, codes of practice and human intuition are not sufficient to form an optimal evacuation plan. Rather attention needs to be paid to mathematical optimization approaches \citep{haghani2020optimising}.

In this paper, we are interested in what will happen in a complex building containing a large crowd, after the sound of a whistle or a fire alarm, in a serious situation, when there is enough yellow-coated rescue personnel, whose members have enough skill and authority to guide the crowd out of the building in minimum time. Even more so, how should these guides be coordinated so that they manage in their task in an optimal way. To our knowledge, this problem has not been solved using a rigorous mathematical optimization approach.

In the research field of crowd and evacuation dynamics, three distinct research streams can be distinguished: experimental research on pedestrian behavior, descriptive mathematical modeling of crowd movement and interactions, and mathematical optimization of evacuation. A good source of references on crowd movement models and optimization approaches is \citep{vermuyten}. The most popular crowd movement model due to its ability to produce realistic physical movement is the microscopic agent-based social force model \citep{socialforce95}. In it the movement of individual agents are described by Newton's equations of motion.

A typical concern is that focusing mainly on physical movement in crowd evacuation modeling may potentially lead to an underestimation of the evacuation time, since evacuees can engage in a variety of activities that do not move them towards the exits \citep{gwynne2016}. We assume, that in an evacuation of a large crowd controlled by guides, such activities can be important but rare. From a modeling perspective, these behavioral deviations are small compared to the systemic, deterministic part of motion \citep{helbing2013}. Deviations from the usual rules of motion can be approximated by adding a Gaussian noise term to the equations of motion \citep{socialforce95}. So, in this study, our main focus is not to tackle various human behaviors, but rather to consider how the guides should be positioned and what kind of routes they should take in a minimum time evacuation, provided some simple assumptions about the crowd behavior in such a situation are made.

There is a need for mathematical approaches for evacuation optimization \citep{kurdi2018effect, haghani2020optimising}. Typically, the optimal guided evacuation has been studied by comparative analysis, where the evacuation is simulated and compared for different guide configurations. The optimal number of guides has been studied in, e.g., \citep{trainedleader, evacuationassistants, optimizingproportion}. Also, initial positioning of guides has been studied in, e.g., \citep{aubeandshield, evacuationassistants, multigridmodel}. It has been suggested that embedded, peripheral, near exit and uniformly distributed positions improve the evacuation efficiency. The importance of coordination has been noted in \citep{leadershipeffect, multigridmodel}. Actually, without strategically planned initial positions and exit assignments of guides, their use can worsen the evacuation instead of improving its efficiency.

To our knowledge, there are only two noteworthy studies that use mathematical optimization to study the evacuation of a crowd using guides \citep{invisiblecontrol, maximumcoverage2}. The lack of research is probably due to the large state space caused by a moving crowd and the stochastic nature of the problem. In \citep{invisiblecontrol} the crowd evacuates from a simple building. Its members are assumed to be unfamiliar with the environment. Hence, their movement is a combination of random walk and following nearby agents. Invisible guides are added to the crowd, i.e., the other agents in the crowd do not recognize them as guides. The minimum evacuation time problem is formulated as a stochastic control problem, where the trajectories of the guides are solved. However, the number of guides is not an optimization variable. \cite{invisiblecontrol} do not directly generate an implementable evacuation plan for rescue guides, but rather do a mathematical study on optimal herding of a crowd to an exit. Still, with small changes, their modeling framework can be used for visible rescue guides. Nevertheless, we are skeptical that the solution methods they propose for the problem would be applicable for a more complicated building geometry with the crowd scattered over it.

The study of \cite{maximumcoverage2} is a simultaneous but independent line of work with ours, and the two works were done without knowledge of each other. \cite{maximumcoverage2} use a deterministic version of the social force model to describe the crowd movement. At the start they optimize the number and initial positions of the guides using a suitable criterion. After that, they optimize their routes and exits given their starting positions. The two optimization problems are independent of each other, and neither the criterion nor the starting positions of the guides are corrected based on the information got from the second problem. It should be noted that in the first problem any suitable criterion, not only that used by \cite{maximumcoverage2}, can be used to fix the number and starting positions of the guides. All they give a different value for the evacuation time; the minimum value can be obtained only by chance. The model of \cite{maximumcoverage2} returns an evacuation plan that does not minimize the evacuation time, but does something else.

Furthermore, both \citep{invisiblecontrol, maximumcoverage2} use their modified versions of the social force model, which are in some sense more sophisticated than typically, but they can only be used for optimization and evacuation simulations in simple building geometries. In them, the follower agents might get lost and are unable to follow the guide, if the guide goes into another room, or even behind a corner ~\mbox{\citep{tracemodel}}.

In this paper, we formulate and solve the problem of minimizing the evacuation time of a crowd from a complex building using rescue guides. The problem is formulated as a stochastic control problem, where the objective is to minimize the expected evacuation time. The optimization variables are the number of guides, and their routes represented as origin-destination pairs. The state of the system is modeled with social force equations and simple rules of interaction between guides and exiting agents. The resulting problem solution space is very large, and it cannot be solved with derivative-based optimization. Thus, we first reformulate the problem as a scenario optimization problem \citep{scenarioapproach}, where the expected evacuation time is replaced by its sampled mean. Then we apply a hybrid numerical simulation and genetic algorithm (GA) approach \citep{goldberg}. In it, the GA iteratively searches for the optimal solution, and numerical simulations evaluate their sample mean evacuation time and steers the randomized search process. We ensure the accuracy and efficiency of the algorithm on a test case, and then apply it on a more complex conference building case. Our three main contributions are that our method applies for complex buildings, it takes into account stochasticity and gives the number of guides, their initial positions and exit assignments needed to minimize the crowd evacuation time in a single optimization.

The paper is structured as follows. In Sec.~\ref{sec:dynamics}, we discuss the assumptions and mathematical details of the crowd movement model. In Sec.~\ref{sec:optmodel}, we formulate the optimization problem, and present the GA approach used to solve it. The performance of the GA is analyzed on a test case in Sec.~\ref{subsec:hexagon}, and after that the GA is applied on a more complex conference building case in Sec.~\ref{sec:conference}. Then, in Sec.~\ref{sec:behavioral}, we revisit and discuss the behavioral assumptions made. Sec.~\ref{sec:implementation} is for implementation details and performance of our algorithm. Finally, Sec.~\ref{sec: conclusion} is for conclusion. For more information about parameter values and exact mathematical expressions of the social force model, see Appendix A. And, for a detailed analysis on the effect of stochasticity on our problem, see Appendix B.

\section{Evacuation model with guides}\label{sec:dynamics}

The social force model was first presented by \mbox{\cite{socialforce95}}. In it the motion of a pedestrian is described by Newton's equations of motion. The model is based on a paradigm from social theory, where an action taken by a pedestrian is understood as a measure of her personal motivation to perform that action. That action is taken according to her psychological processes of assessment of alternatives and utility maximization. In the case of pedestrian behavior, this motivation evokes the physical production of an acceleration force. One can say that a pedestrian acts as if she would be subject to external forces (Helbing \& Moln\'{a}r, 1995; Hoogendoorn \& Bovy, 2003; Helbing \& Johansson, 2013).
Additional physical interaction forces, inspired from driven granular flows, are assumed to come into play, when pedestrians get so close to each other that they have physical contact.
\nocite{differentialgames}

The social force model has been integrated into the Fire Dynamics Simulator with Evacuation (FDS+Evac) of the National Institute of Standards and Technology \mbox{\citep{nist}}. We will use the version of the social force model found in the FDS+Evac user manual \mbox{\citep{fdsevac}}. But instead of approximating the human body with three overlapping circles \mbox{\citep{heliovaara2012counterflow}}, we use a single circle. Also, we make a small modification to the social force term. In the original social force model, a distance-dependent term is assumed. However, \mbox{\cite{powerlaw}} have analyzed a large amount of publicly available crowd data, from several outdoor environments and controlled laboratory settings, showing that the social force depends on the time-to-collision between agents rather than their distance. We will use the social force term presented in \mbox{\citep{powerlaw}}.

\subsection{Detailed description}

The evacuation model describes the evacuation of a crowd of agents $N \cup G$ from a space $\Omega \subset \mathbb{R}^2$ after the alarm has been given. The space includes obstacles $\mathcal{O} \subset \Omega$ and exits $\mathcal{E} \subset \Omega$. The obstacles are line segments, or walls, with index set $W$. The crowd consists of exiting agents $N$ and guide agents, or guides, $G$. Here $N=\left\lbrace 1,...,n \right\rbrace $ and $G= \left\lbrace n+1,...,n+m \right\rbrace $ are the index sets of the exiting agents and guides, respectively. Exiting agents represent regular people that do not have full knowledge of the exits in the building, and head to their familiar exits by default. The guides represent trained safety personnel that use routes instructed by the evacuation planner. They are assumed to have enough authority to influence the route choices of exiting agents. More specifically, the first time an exiting agent is within the interaction range $r^{guide}$ from a guide, it receives information of the exit the guide is moving towards, and starts also heading there. However, in a threatening situation, we assume that an exiting agent wants to get out of the building as fast as possible. That is why, if an exit is within the visibility range $r^{exit}$, it starts to head there instead, regardless of where it was previously heading. In all situations, the exiting agents move to the exits using a shortest path (for information on its numerical computation see \mbox{Sec.~\ref{sec:implementation}}).

It is assumed that a mixture of socio-psychological and physical forces influence the agents' motion in the crowd. At time $t$, agent $i \in N \cup G$ with mass $m_i$ and radius $r_i$ likes to move with a certain desired velocity $\mathbf{v}_i^{des} (t)$, where $\mathbf{v}_i^{des} (t) = v_i^{des} (t) \mathbf{e}_i^{des} (t)$. Here, $v_i^{des} (t)$ is its desired speed, and $\mathbf{e}_i^{des} (t)$ is a unit vector that points to the direction that gives the shortest path to the exit it is heading towards. Agent $i$ attempts to change its actual velocity $\mathbf{v}_i (t)$ to $\mathbf{v}_i^{des} (t)$ with a reaction time $\tau^{reac}$.

If agents $i$ and $j$, $j \in N \cup G, j \neq i$, are about to collide, they experience a repulsive social force $\mathbf{f}_{ij}^{soc} (t)$. When agent $i$ is in contact with another agent $j$ or wall $w \in W$, the physical contact forces $\mathbf{f}_{ij}^c (t)$ or $\mathbf{f}_{iw}^c (t)$, respectively, arise. Additionally, we assume that agent $i$ is affected by a small random fluctuation force $\bm{\xi}_i(t)$. Typically, the social force model includes a random force term in each agent's equation of motion. This force represents random deviations in behavior; more precisely, situations where two or more behavioral alternatives are equivalent, i.e., whether to pass an obstacle from the left or right hand side. It can also be thought to describe accidental or intentional deviations from the usual rules of motion \citep{socialforce95}. The force $\bm{\xi}_i(t)$ is drawn from a truncated bivariate normal distribution with zero $\bm{0}$ mean. However, for guides this force equals a zero vector $\mathbf{0}$. For the parameter values used and the exact mathematical expressions of the forces, see Appendix A.

The change of velocity at time $t$ for agent $i$ is then given by the equation of motion:
\begin{equation}
m_i \dfrac{d\mathbf{v}_i}{dt} = m_i \dfrac{\mathbf{v}_i^{des} - \mathbf{v}_i}{\tau^{reac}}  + \sum_{j(\neq i)} (\mathbf{f}_{ij}^{soc} + \mathbf{f}_{ij}^{c}) + \sum_{w} \mathbf{f}_{iw}^{c} + \bm{\xi}_i;
\label{eq:eqofmotion}
\end{equation}

\noindent
the change of position vector $\mathbf{x}_i (t)$ is given by the velocity
\begin{equation}
\dfrac{d\mathbf{x}_i(t)}{dt}=\mathbf{v}_i(t).
\label{eq:velocity}
\end{equation}

\noindent
In addition, we assume the following interactions to take place in the evacuation:

\begin{description}[align=left, style=unboxed, leftmargin=0cm]
\item [Assumption 1.] An exiting agent heads to its familiar exit by default.
\item [Assumption 2.] A guide moves from its starting position to its target exit using the shortest path.
\item [Assumption 3.] If an exiting agent is moving towards its familiar exit, and a guide comes within the interaction range $r^{guide}$, it starts to move to the same exit as the guide. If multiple guides are within the $r^{guide}$ range, it starts to follow the closest one.
\item [Assumption 4.] If an exiting agent follows a guide, it will not switch to follow another guide.
\item [Assumption 5.] If an exit is within the visibility range $r^{exit}$ from an exiting agent, it heads there, whether it previously was following a guide or heading to its familiar exit. If multiple exits are within the $r^{exit}$ range, it heads to the closest one.
\end{description}

\section{Optimization model}\label{sec:optmodel}

We are interested in minimizing the evacuation time of the crowd, or equivalently the evacuation time of the last evacuated agent $T_{last}$. If we denote the evacuation times of the agents in the crowd with $t_1,...,t_{n+m}$, the maximal element of the set is $T_{last}$. Because of the random fluctuation force term $\bm{\xi}_i (t)$ in Eq.~\eqref{eq:eqofmotion}, our problem is stochastic. Thus, as objective function we choose to minimize the expected value of $T_{last}$ with respect to $\bm{\xi}_i(t), t \in [0, T_{last}], 1 \leq i \leq n$. Recall that for $i \in G$ the random fluctuation force term is equal to $\mathbf{0}$.

We discretize the space $\Omega$ into square grid cells $\omega$, so that $\Omega \subset \cup \omega = \bar{\Omega}$. Also, we denote the points associated with an exit by $\varepsilon \subset \mathcal{E}$, so that $\mathcal{E} = \cup \varepsilon$. Each guide $g \in G =\{n+1,...,n+m\}$ is associated with a starting grid cell $\omega_g  \subset \bar{\Omega}$, and a target exit $\varepsilon_g \subset \mathcal{E}$. The optimization variables are $(\omega_g, \varepsilon_g), g \in G$. We assume that the number $m$ of the possible guides is sufficiently large, so that one or more guides can remain idle in the simulation. The idle guides are mapped to a dummy grid cell outside $\bar{\Omega}$, and their evacuation time is defined to be zero. Our focus here is either a fixed number of active guides used in optimization, or a set of active guides obtained as a result of optimization. 

The initial position of a guide $g$, $\mathbf{x}_g (0) \in \omega_g$, is a prespecified point in its corresponding starting grid cell $\omega_g$. The end position of a guide $g$, $\mathbf{x}_g (t_g) \in \varepsilon_g$, is any point in its corresponding target exit $\varepsilon_g$. The initial positions of the exiting agents are prespecified $\mathbf{x}_i(0)=\mathbf{x}_{i,0}, i=1,...,n$. The exiting agents can evacuate using any of the available exits; thus it holds for the end positions $\mathbf{x}_i (t_i) \in \mathcal{E}$.

The agents move to the exits according to Eqs.~\eqref{eq:eqofmotion},~\eqref{eq:velocity} and Assumptions 1-5 defined in Sec.~\ref{sec:dynamics}. These equations constitute the constraint equations of this problem. Now we can formulate the problem of minimizing the evacuation time of a crowd using rescue guides:
\begin{align}
& \min_{({\omega}_g, {\varepsilon}_g)} \quad \mathbb{E} \big[ T_{last} | \bm{\xi}_i (t), 1 \leq i \leq n, t \in [0, T_{last}] \big]; \nonumber \\
& {\omega}_g \subset \bar{\Omega}, {\varepsilon}_g \subset \mathcal{E}, g \in G \nonumber \\
& \label{eq:dynamicopt} \\
& \textrm{subject to Eqs.}~\eqref{eq:eqofmotion},~\eqref{eq:velocity}; \textrm{Assumptions 1-5}; \nonumber \\
& \mathbf{x}_i (0)=\mathbf{x}_{i,0}, \mathbf{x}_i(t_i) \in \mathcal{E}, i \in N; \mathbf{x}_g(0) \in \omega_g, \mathbf{x}_g (t_g) \in \varepsilon_g, g \in G. \nonumber
\end{align}

To solve problem \mbox{Eq.~\eqref{eq:dynamicopt}} we will use scenario optimization \mbox{\citep{scenarioapproach}}. The expected evacuation time of the last agent is replaced by its sampled empirical version: $M$ independent identically distributed samples of the random force $\bm{\xi}_i, 1 \leq i \leq n$, are generated which we denote by $\bm{\delta}^{(1)},...,\bm{\delta}^{(M)}$, and then we define the sample mean,
\begin{equation}
\hat{T}_{last}=\dfrac{1}{M} \sum_{k=1}^M T_{last}^{(k)}.
\label{eq:empiricalmean}
\end{equation}

Here, $T_{last}^{(k)}$ is calculated for given (${\omega}_g$, ${\varepsilon}_g$), $g \in G$, by numerically simulating the system equations \mbox{Eqs.~\eqref{eq:eqofmotion},~\eqref{eq:velocity}} with a numerical integration scheme, given the sample vector $\bm{\delta}^{(k)}$, Assumptions 1-5 and the initial and end conditions (see more implementation details in \mbox{Sec.~\ref{sec:implementation}}).

The scenario optimization problem is solved with a genetic algorithm (GA) \citep{goldberg}. The GA iteratively searches for the optimal solution, while the numerical simulations evaluate the fitness of the found solutions and steer the randomized search process. In our optimization problem, the fitness is the sample mean Eq.~\eqref{eq:empiricalmean}, and a solution, or chromosome, is an evacuation plan $\left\{ (\omega_{n+1}, \varepsilon_{n+1}),...,(\omega_{n+m}, \varepsilon_{n+m}) \right\}$. A single gene in a chromosome contains both the starting grid cell and target exit of a guide. Because a solution has a variable number of active guides, we apply the hidden genes GA \citep{hiddengene}. In it each gene gets a tag, which tells if the gene is active, or if it is hidden or idle. Only the active genes affect the evacuation simulation. However, the genetic information of the idle genes is carried on in the process.

In the GA, a population of a predefined number of solutions is maintained in the consecutive iterations, or generations. At the initial iteration, a population of random solutions is generated. After that, the population goes through three operations, which are selection, crossover and mutation. In selection, the solutions with smallest fitness values are chosen to undergo the next two operations. In this paper, we use elitist selection, where we replace a few of the worst solutions in the current generation with the best solutions of the previous generation.
 
\begin{figure}[ht!]
\centering\includegraphics[width=0.7\textwidth]{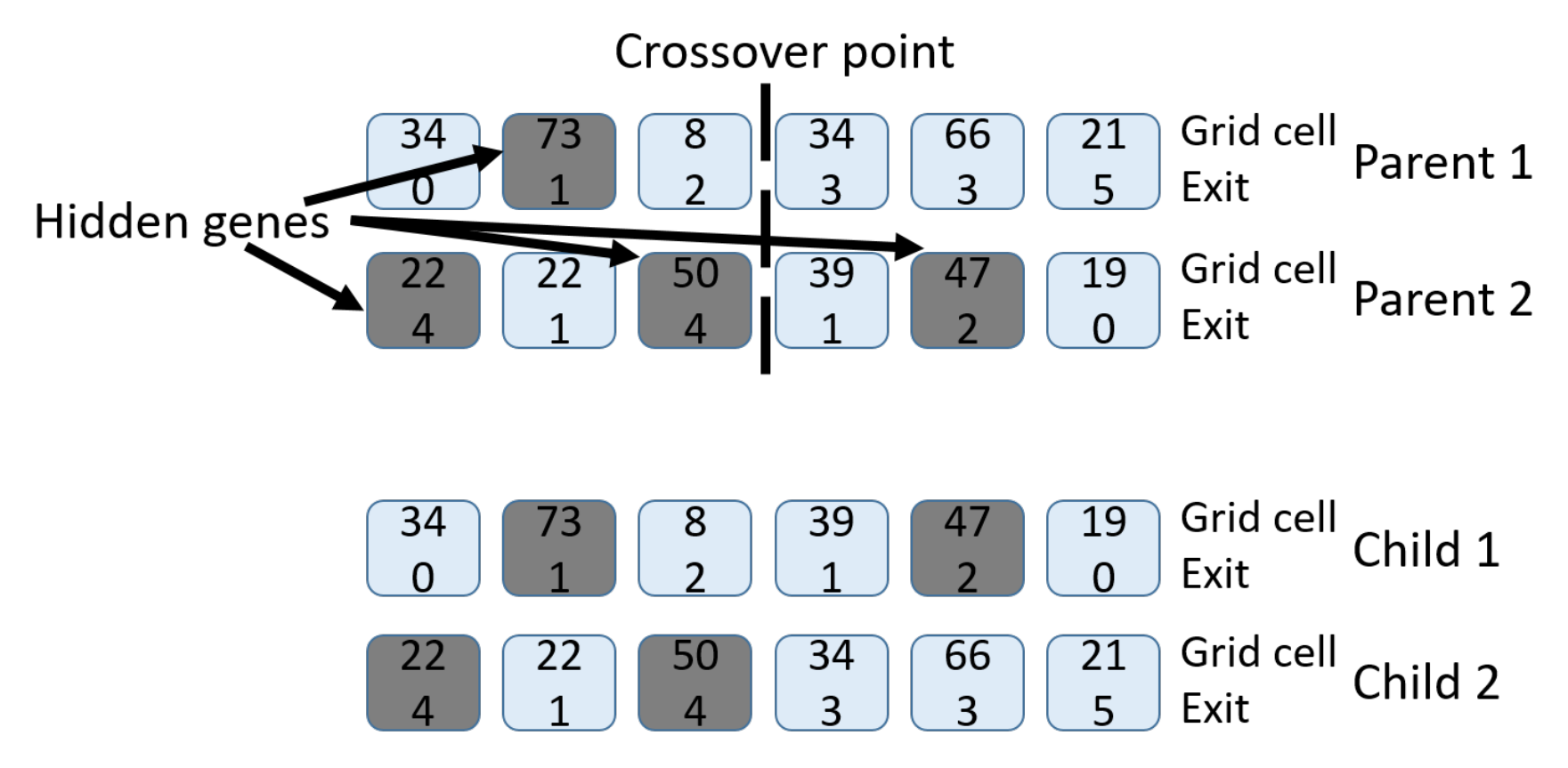}
     \caption{The single-point crossover operation applied on the two parent chromosomes to create offspring solutions.}
  \label{fig:figure_1}
\end{figure}

In the crossover operation, genetic information is combined to create new solutions. In this paper, we use the single-point crossover operator, meaning that the offspring solutions of the two parent solutions contain half of the genetic material of both parents. The crossover operator is applied with a certain probability, and it is applied on both genes and the tags. If it is not applied, the offspring solutions have exactly the same genetic information as their parents. See Fig.~\ref{fig:figure_1} for a depiction of the mechanism; if the gene is colored gray, the gene is hidden, which means that the guide corresponding to the gene is not present in the evacuation, i.e., is idle. In the single-point crossover, the genetic material on the right side of the crossover point of the two parent chromosomes is changing places to create offspring chromosomes.

Then, a mutation operator is applied on the offspring solutions. The mutation operator is applied separately, with a certain probability, on each gene and tag. When applied on a gene, it can either alter the starting grid cell, or the target exit, or both of them. The mutation operator applied on a tag can switch the gene from hidden to active, or vice versa. Finally, the GA goes back to evaluating the fitness values of the solutions of the new generation, i.e., the offspring solutions, and then proceeds to the selection operation, and the iteration continues. The algorithm has converged, when the best solution has not changed in a predefined number of successive iterations.

\section{Evacuation of a hexagon-shaped area}\label{subsec:hexagon}

Typically, to solve an optimization problem with a GA, the algorithm parameters are tuned manually in a problem-specific manner. Hence, we will first apply the GA on a test case, of which we have some idea what the optimal evacuation plan might be. The algorithm parameters are tuned, so that it converges as efficiently and accurately as possible to the near-optimal solution. The same parameters are also used for the more complex case presented in the next section.

Here, we consider a hexagon-shaped area. We assume it to be an outdoor space with six exits. All the exiting agents have entered through one door, and are thus familiar only with that one. Also, we assume this to be an unhurried situation, where people want to move out of the area, but do not sense such an urgency that they would by themselves try other exits. Thus, we do not use Assumption 5. The interaction range of a guide is set to $r^{guide}=5$ m.

For exiting agents $i \in N$, the initial positions $\mathbf{x}_i^0$, radii $r_i$, masses $m_i$, and desired speeds $v_i^{des}$ are fixed for all simulations. Before fixing the values, the parameters $m_i$, $r_i$, and $v_i^{des}$ are drawn from a truncated normal distribution with a cutoff at three times the standard deviation. The mean and standard deviations are $73.5$ kg and $8.0$ kg, $0.255$ m and $0.035$ m, and $1.25$ m/s and $0.3$ m/s, respectively for $m_i$, $r_i$ and $v_i^{des}$. On the other hand, for guide agents $g \in G$, we set the values of a typical male: $m_g=80$ kg, $r_g=0.27$ m, and $v_g^{des}=1.15$ m/s. The reaction time is $\tau^{reac}=0.5$ s. These values are all taken from the FDS+Evac user manual \mbox{\citep{fdsevac}, and will also be used for the more complex case in the next section.}

\begin{figure}[ht!]
\centering
\subfigure[][]{
\label{fig:figure_2a}
\includegraphics[height=0.25\textheight]{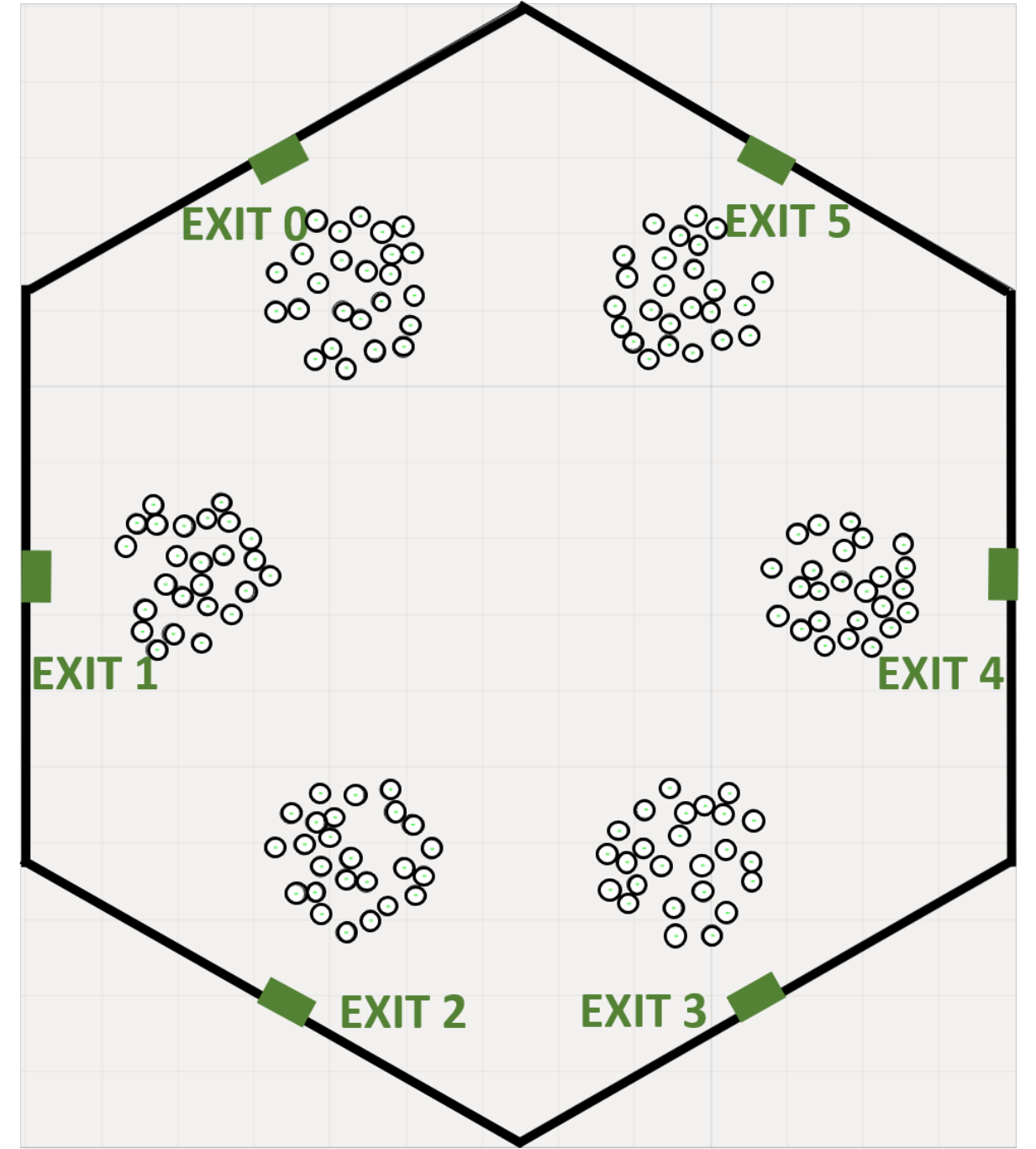}}
\hspace{8pt}
\subfigure[][]{
\label{fig:figure_2b}
\includegraphics[height=0.25\textheight]{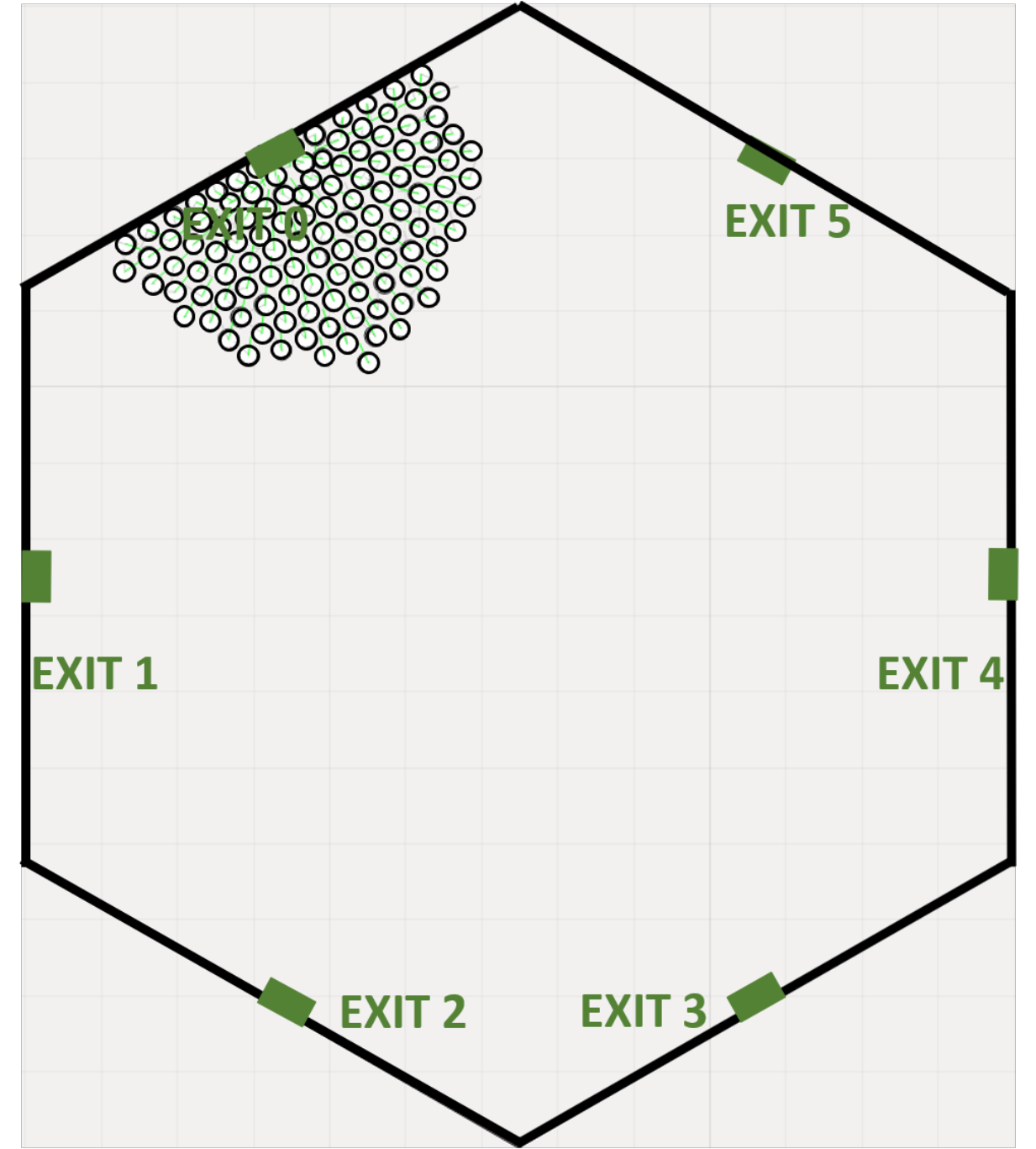}}

\subfigure[][]{
\label{fig:figure_2c}
\includegraphics[height=0.25\textheight]{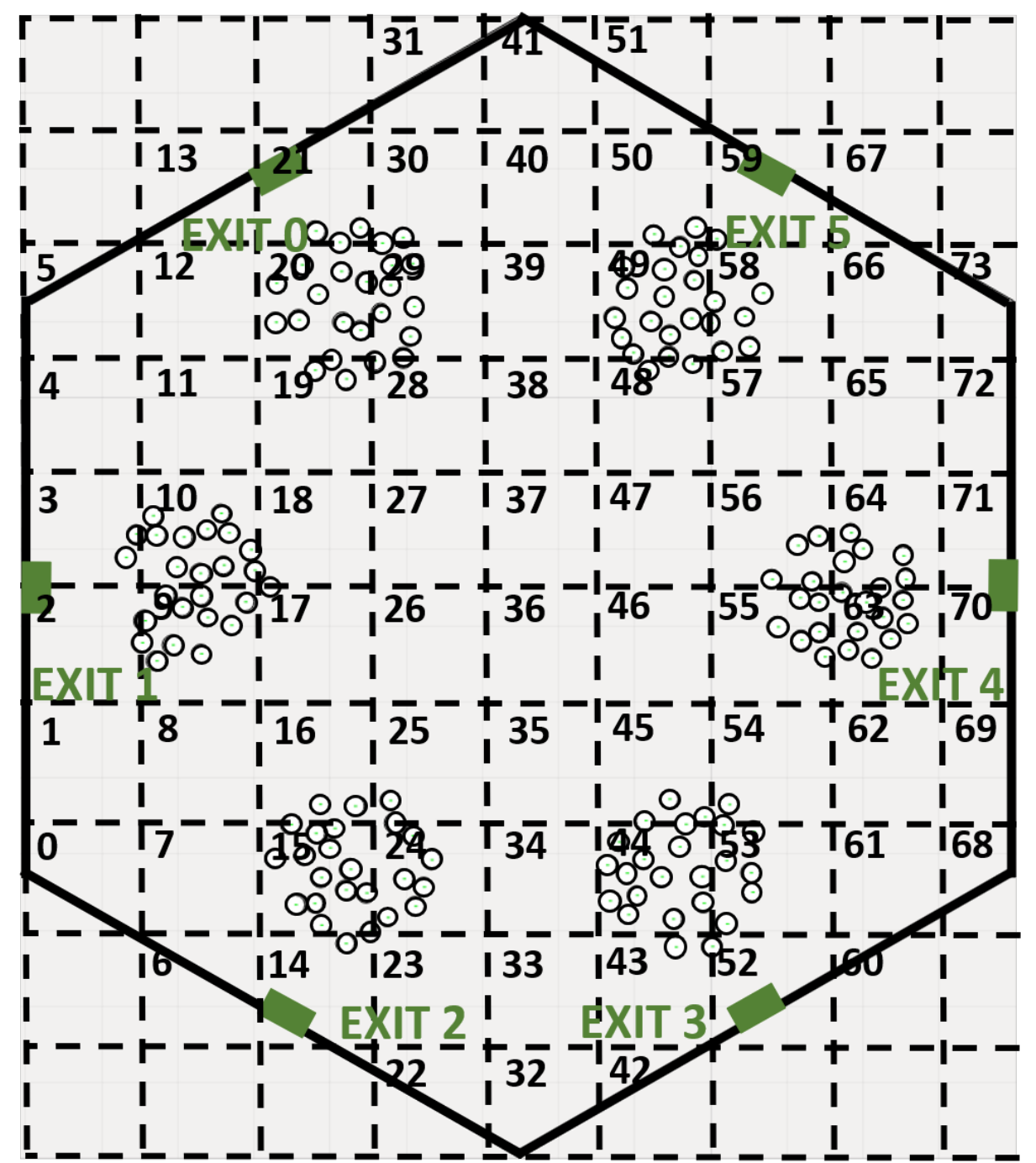}}
\hspace{8pt}
\subfigure[][]{
\label{fig:figure_2d}
\includegraphics[height=0.25\textheight]{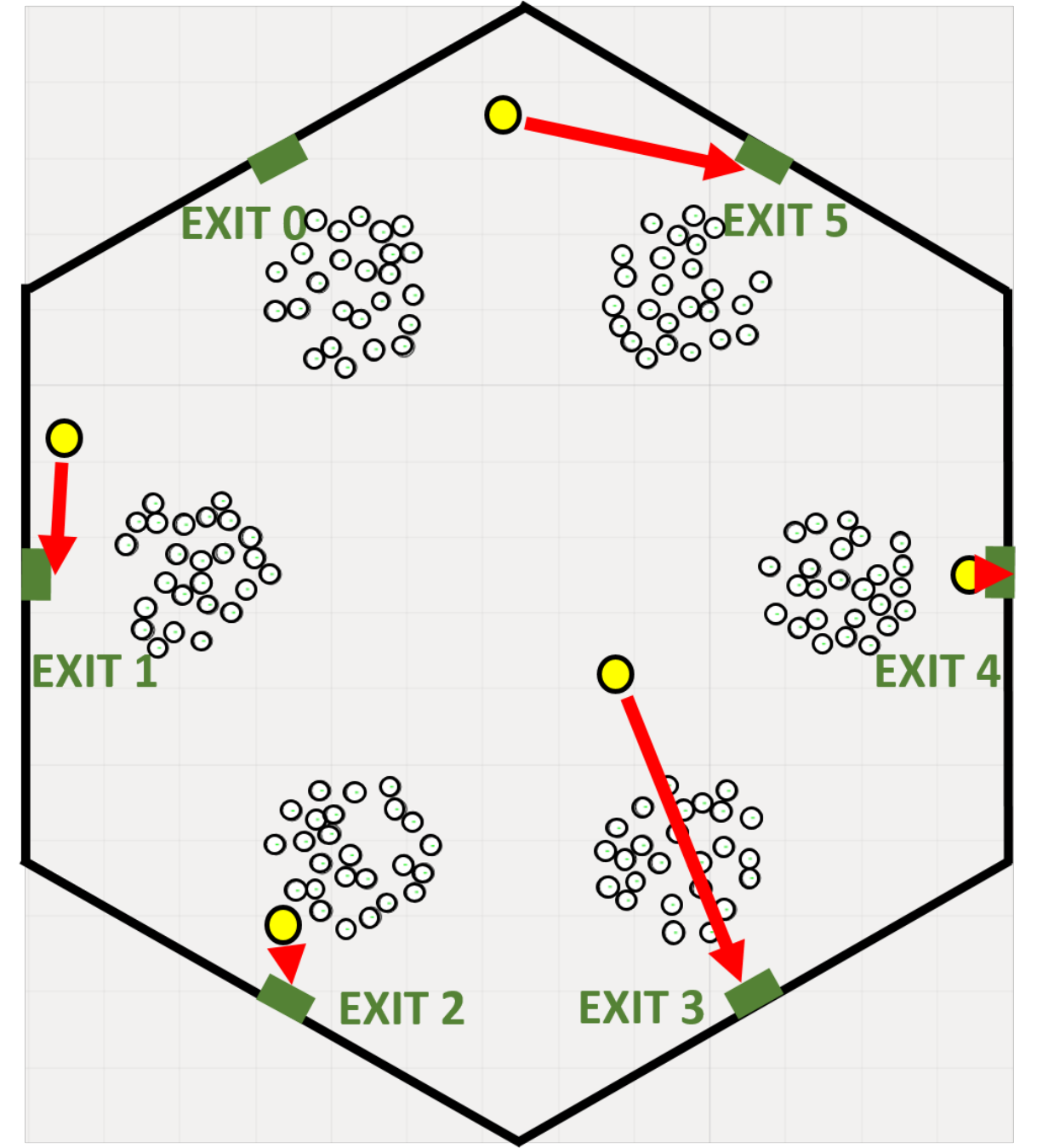}}
\caption[A set of four subfigures.]{Evacuation of a hexagon-shape area. \mbox{\subref{fig:figure_2a}} Initial situation. \mbox{\subref{fig:figure_2b}} Without guides, a jam forms after a while. \mbox{\subref{fig:figure_2c}} Feasible starting grid cells for guides. \mbox{\subref{fig:figure_2d}} The near-optimal evacuation plan.}
\label{fig:hexagon}
\end{figure}

At the beginning of the evacuation, there are six groups of agents, each of them containing $25$ agents, and each of them are located close to an edge of the hexagon; see Fig.~\ref{fig:figure_2a}. When the agents start to evacuate, they all head towards the familiar exit (Assumption 1), and form a jam in front of it; see Fig.~\ref{fig:figure_2b}.

The guides are added to the crowd by applying the hidden genes GA. First, we discretize the space into seventy-four $3$ m $\times$ $3$ m grid cells, which are the feasible starting grid cells for the guides; see Fig.~\ref{fig:figure_2c}. The six exits in the area are the feasible target exits. Because there are only six exits, and the space is highly regular, we expect that no more than six guides should be needed to obtain the optimal evacuation plan, thus a chromosome in the GA contains at most six active genes. A good guess is that the optimal solution most likely utilizes all six exits evenly. Thus, the evacuation time of the optimal solution should be close to the evacuation that utilizes evenly all six exits. Thus, the algorithm parameters are tuned so as to reach that solution as efficiently and accurately as possible. The crossover probability is set to $0.85$, the mutation probability to $0.10$, population size to $40$, with $30$ samples of each chromosome. The two worst individuals are always replaced with two best individuals of the previous generation. The algorithm is considered to have converged to a near-optimal solution, when the best solution has not changed for $15$ generations.

The near-optimal solution of the GA is depicted in Fig.~\ref{fig:figure_2d}. It includes only five guides (depicted by the yellow circles). The guides head from their initial positions to their prespecified exits along the shortest paths depicted by the red arrows (Assumption 2). The exiting agents start to follow their closest guides, within the $r^{guide}$ range, and do not switch to follow another guide (Assumptions 3 and 4). As it was mentioned earlier, all of the exiting agents have already a single familiar exit, so it is enough to redirect the majority of the exiting agents to the other five exits, thus utilizing all six exits in the evacuation.

\begin{figure}[ht!]
 \centering\includegraphics[width=0.5\textwidth]{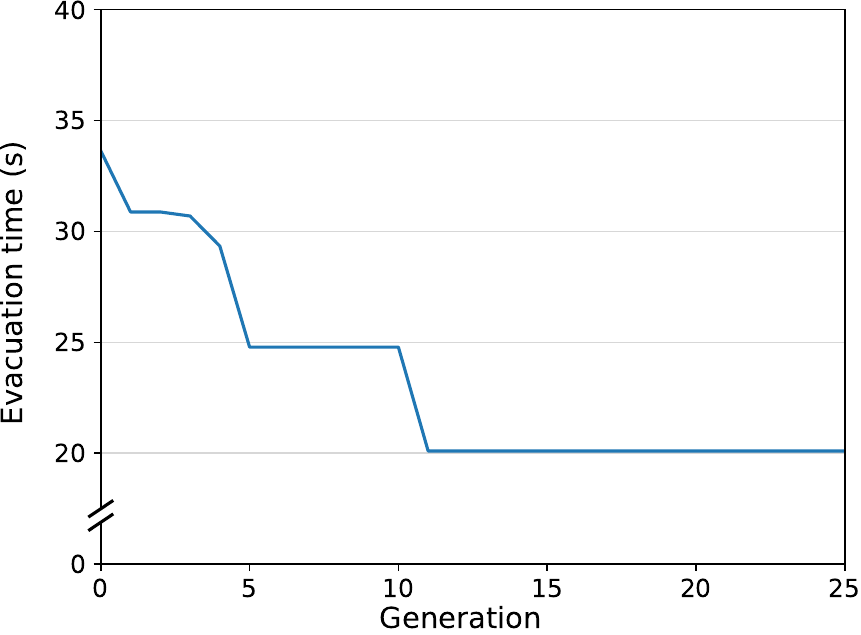}
     \caption{The sample mean evacuation time for the best solution of each generation of the GA, when applied to the test case.}
  \label{fig:figure_3}
\end{figure}

One could have expected that the guides would be in symmetric positions close to their target exits. It is not needed, because the evacuation time is affected both by the walking and the queuing time of the agents. Additionally, the flow at the exit is known to be a nonlinear function of the size of the crowd in front of it \mbox{\citep{schadschneider}}. Thus, up to a certain point, the flow can be increased by increasing the size of the crowd in front of the exit. However, after a certain point the flow will start to decrease. Due to these effects, the near-optimal solution is not very sensitive to the starting positions of the guides, as long as the exit utilization is fairly even, and each guide influences about $1/6$ of the exiting agents. If the interaction range $r^{guide}$ would be increased, the positions of the guides would have to be chosen more carefully, so that the guides do not also influence exiting agents farther away. In Fig.~\ref{fig:figure_3} we see how the hidden genes GA converges to the near-optimal evacuation plan. Starting from the 11th generation, the best solution does not change for 15 generations, and the GA is considered to have converged.

\begin{figure}[ht!]
 \centering\includegraphics[width=0.5\textwidth]{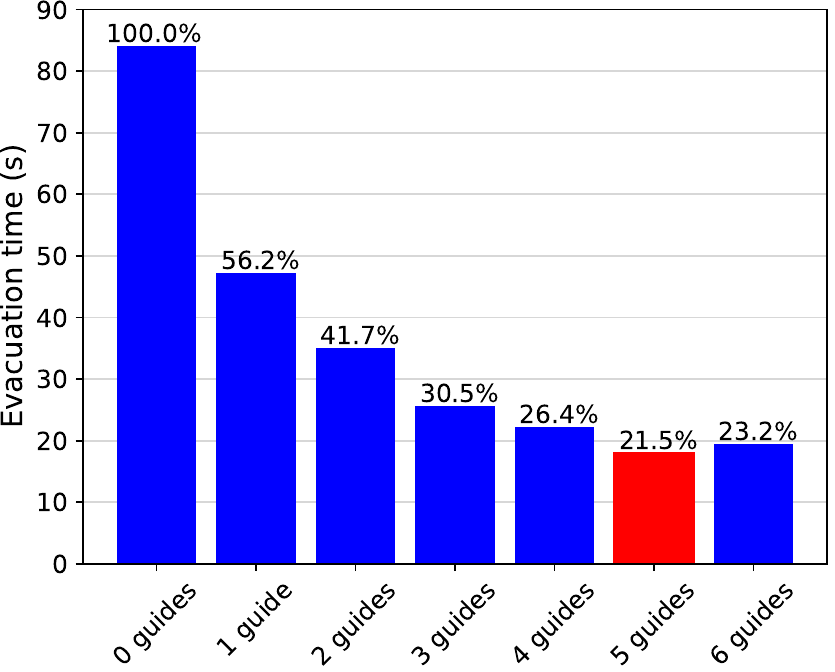}
     \caption{The sample mean evacuation time of the near-optimal evacuation plan for different fixed numbers of guides.}
  \label{fig:figure_4}
\end{figure}

We also try how the solution changes with a fixed number of active guides, i.e., using the GA without hidden genes. We use a larger mutation probability $0.40$ for the $30$ first generations, and after that we change it to only $0.05$. We are able to have a clear exploration phase in the beginning of the algorithm. It is needed for convergence accuracy, when the hidden genes are not used. Otherwise, we use the same settings as with the hidden genes GA. In \mbox{Fig.~\ref{fig:figure_4}} we see a comparison of the sample mean evacuation time for the near-optimal evacuation plans with different numbers of guides. We see that adding guides speeds up the evacuation, up to five guides. The slight increase in evacuation time with six guides is a statistical deviation rather than the sixth guide slowing the evacuation, as we will soon see from the depictions of the evacuation plan. Notice also that the marginal utility of adding guides is decreasing, i.e., adding the first guide cuts the sample mean evacuation time almost in half, but increasing from four to five guides has only a 5 \% benefit. 

\begin{figure}[htb!]
\centering
\subfigure[][]{
\label{fig:figure_5a}
\includegraphics[height=0.25\textheight]{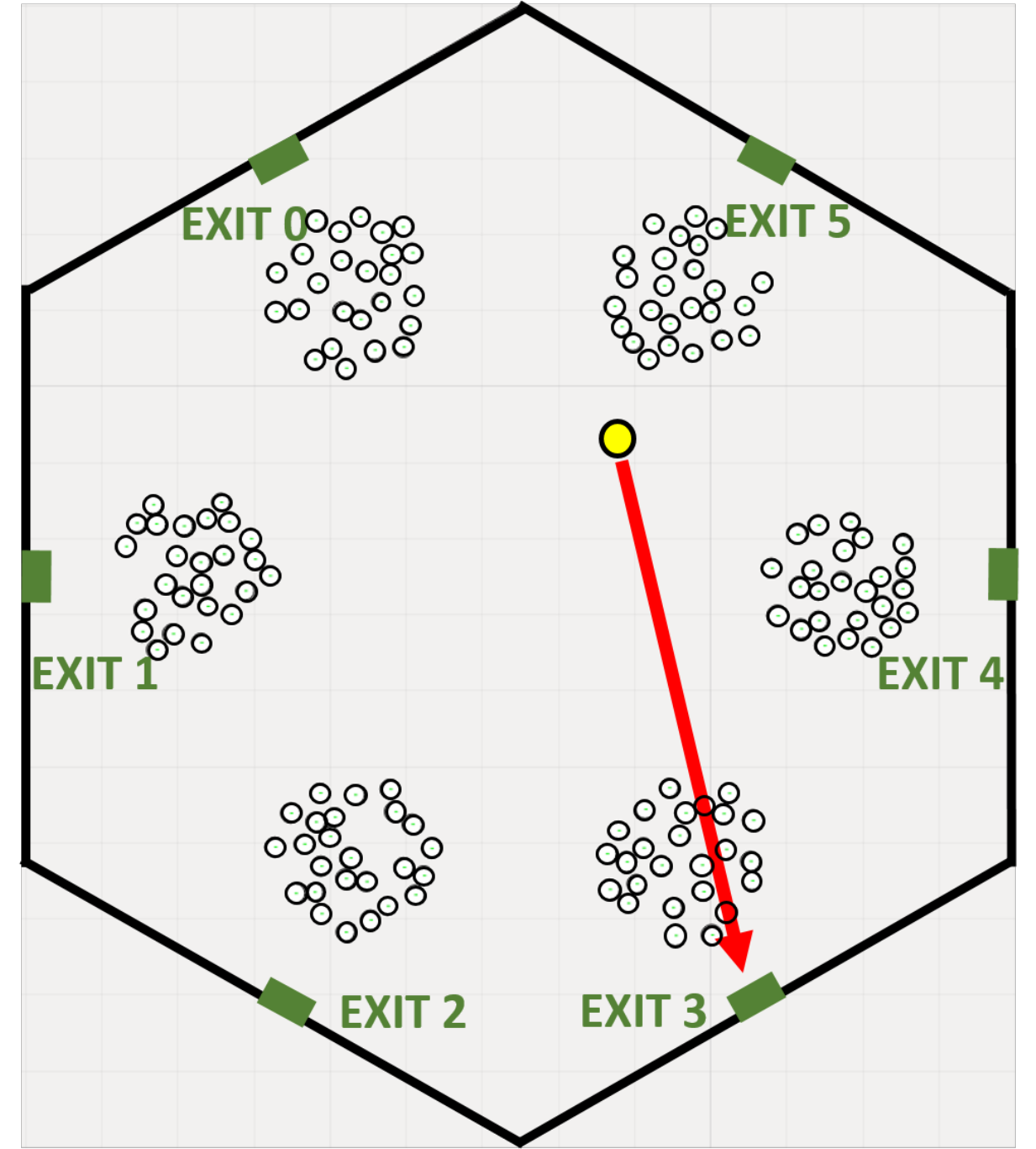}}
\hspace{8pt}
\subfigure[][]{
\label{fig:figure_5b}
\includegraphics[height=0.25\textheight]{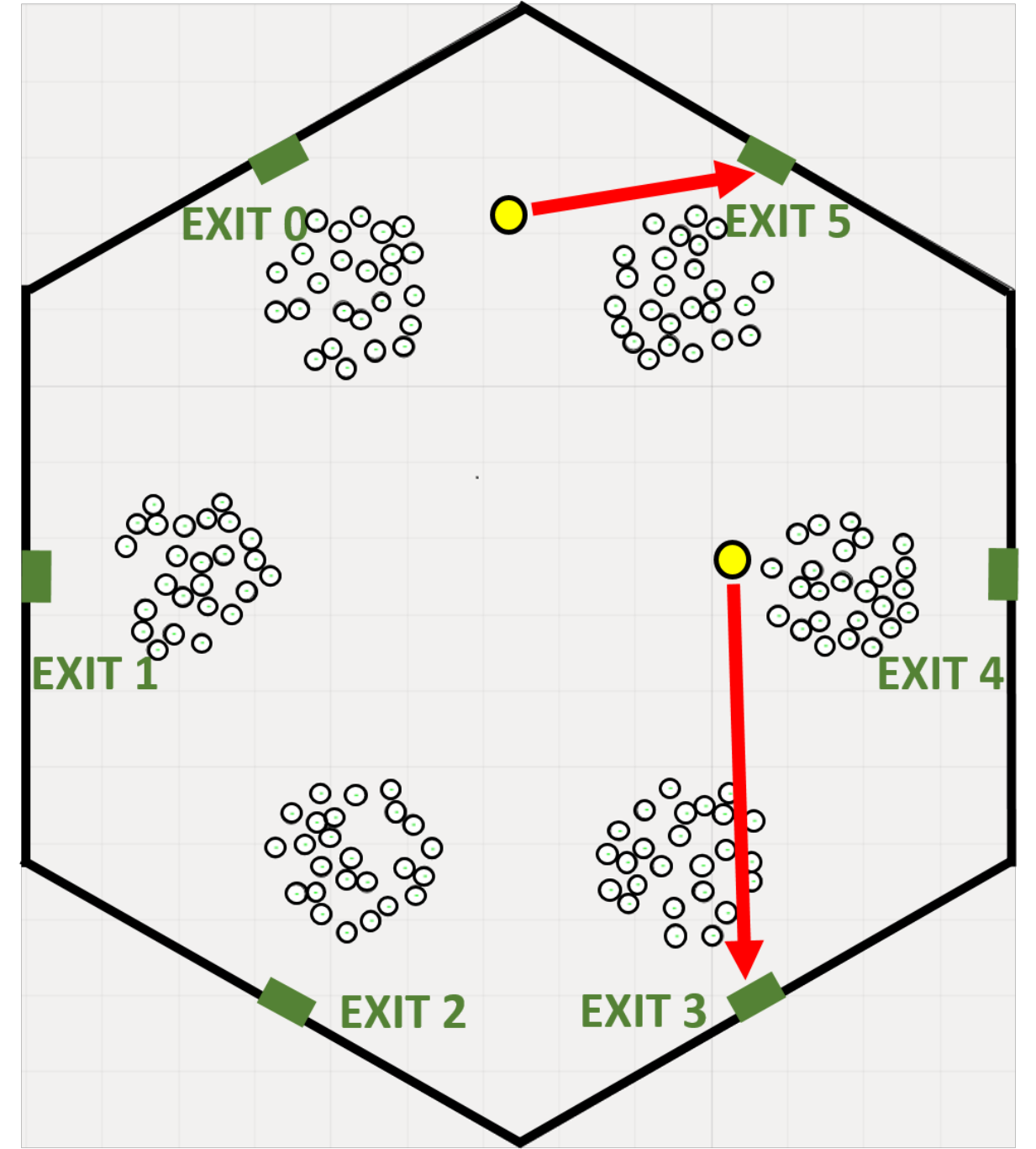}}

\subfigure[][]{
\label{fig:figure_5c}
\includegraphics[height=0.25\textheight]{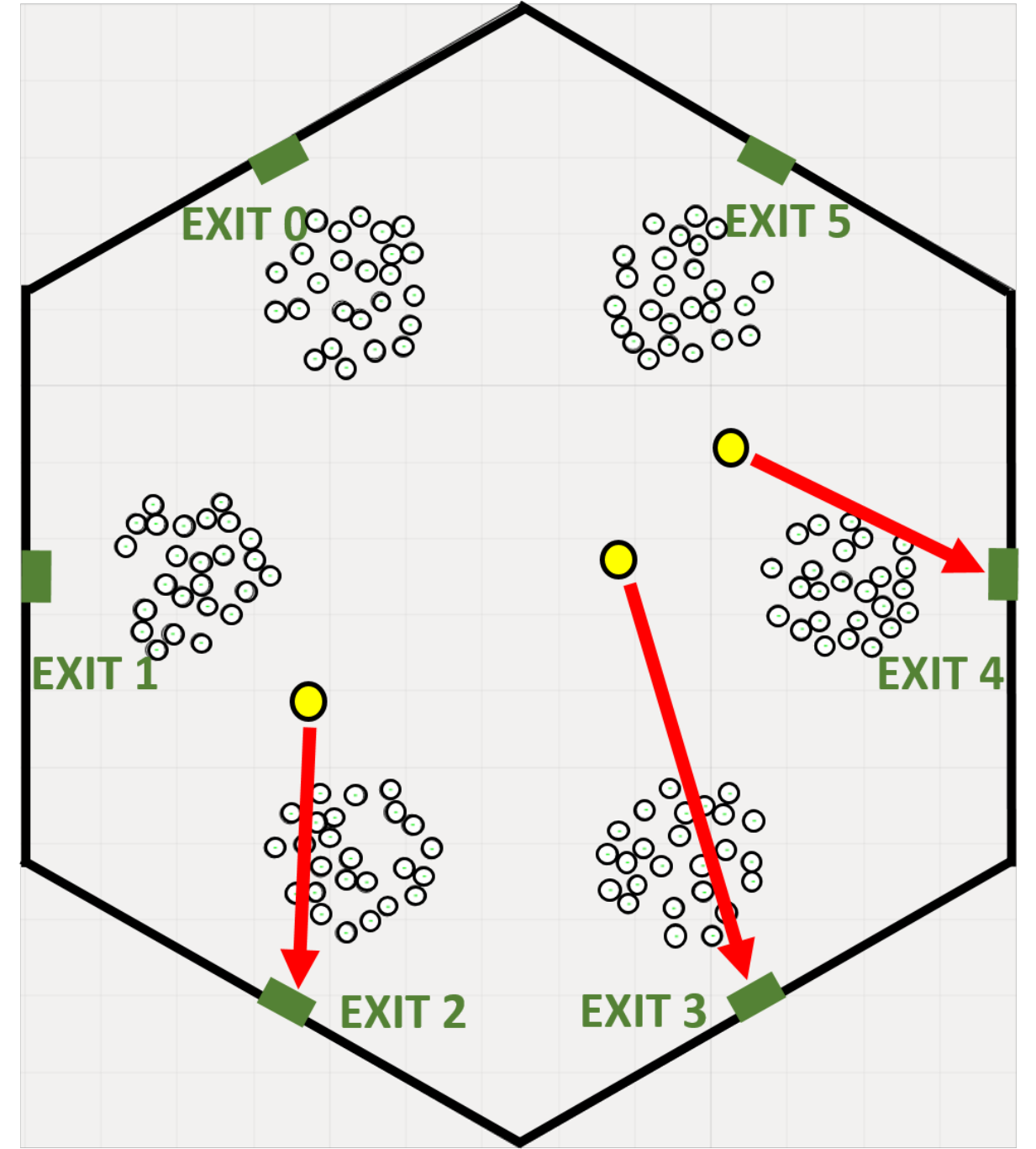}}
\hspace{8pt}
\subfigure[][]{
\label{fig:figure_5d}
\includegraphics[height=0.25\textheight]{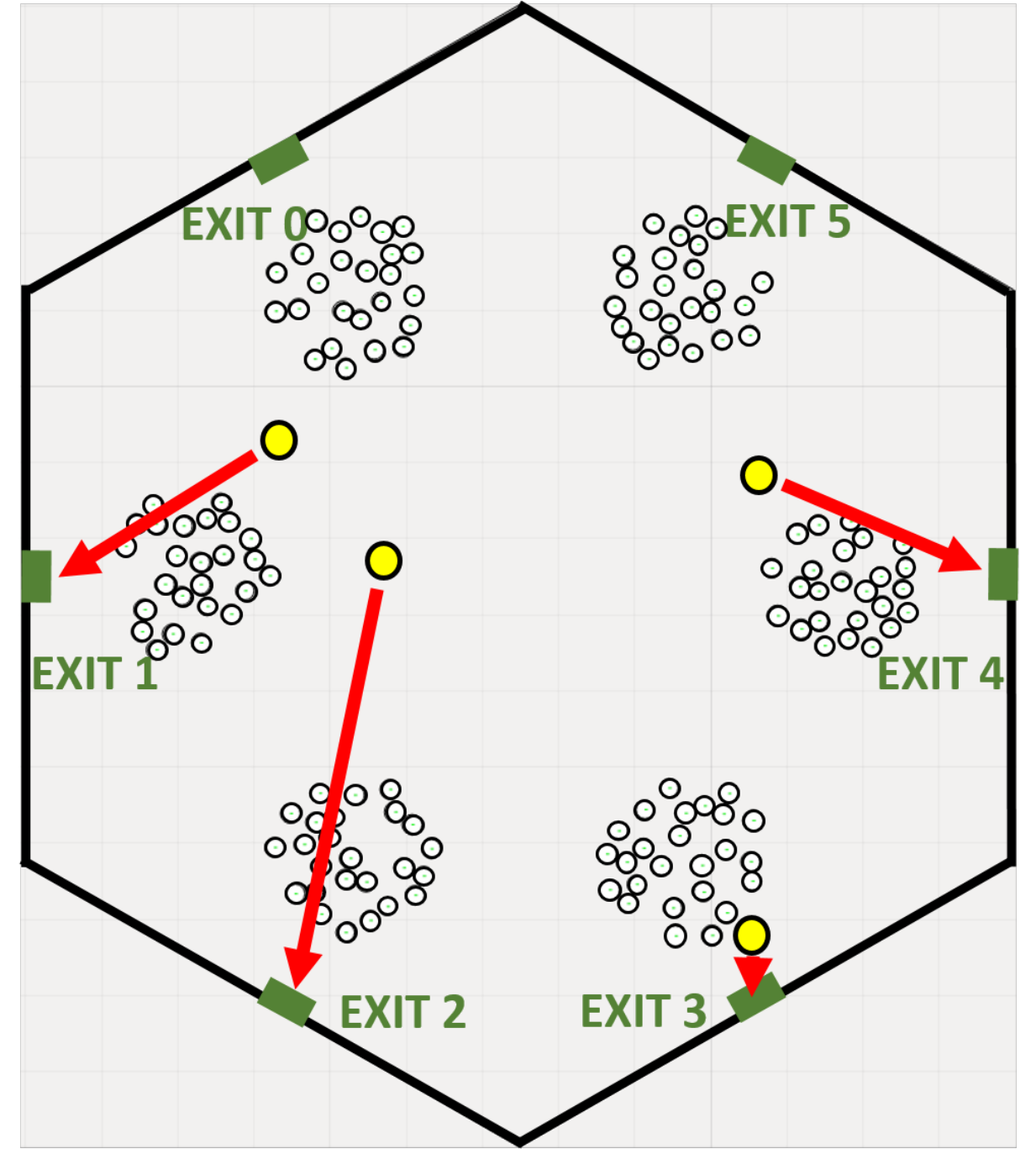}}

\subfigure[][]{
\label{fig:figure_5e}
\includegraphics[height=0.25\textheight]{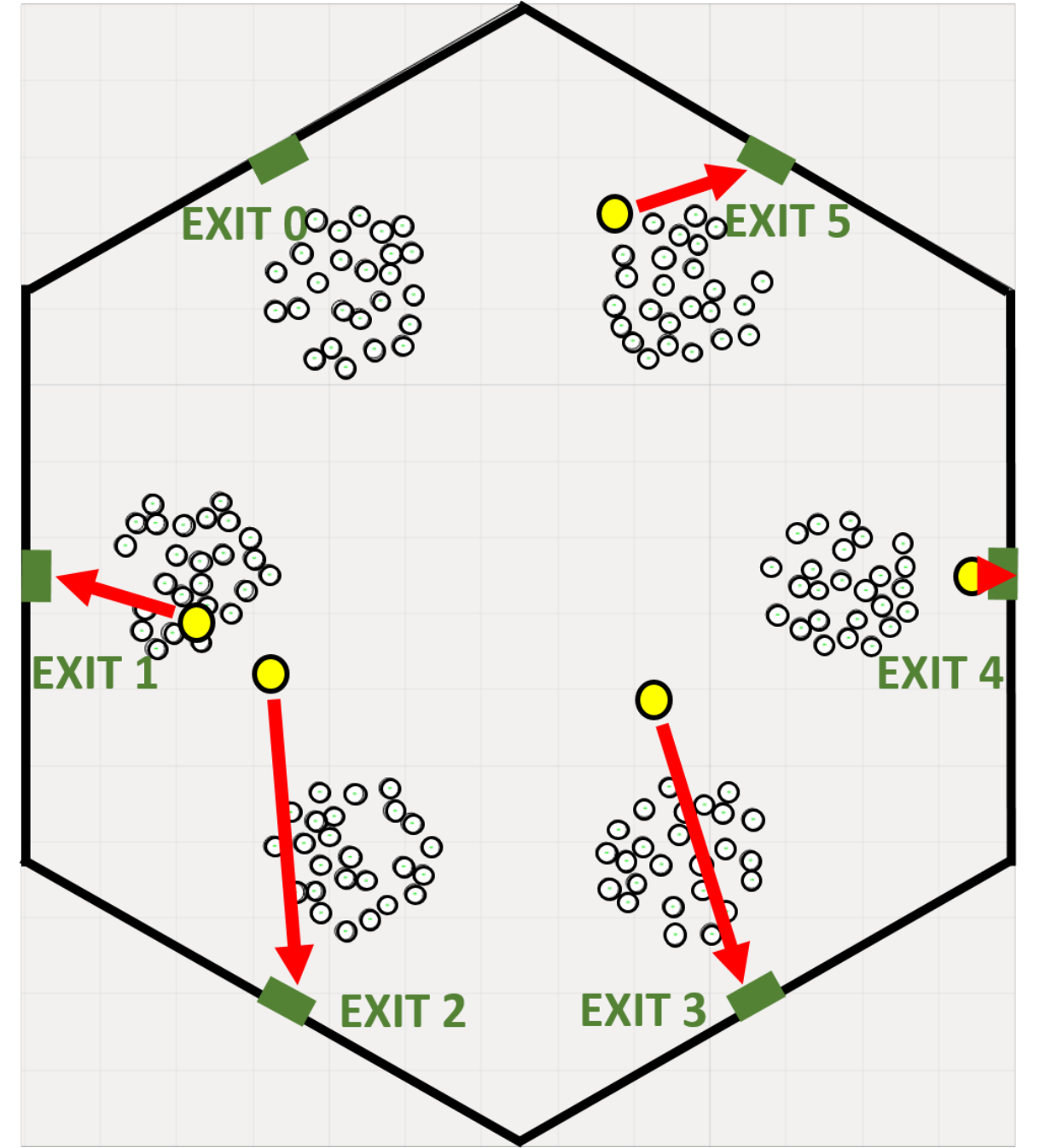}}
\hspace{8pt}
\subfigure[][]{
\label{fig:figure_5f}
\includegraphics[height=0.25\textheight]{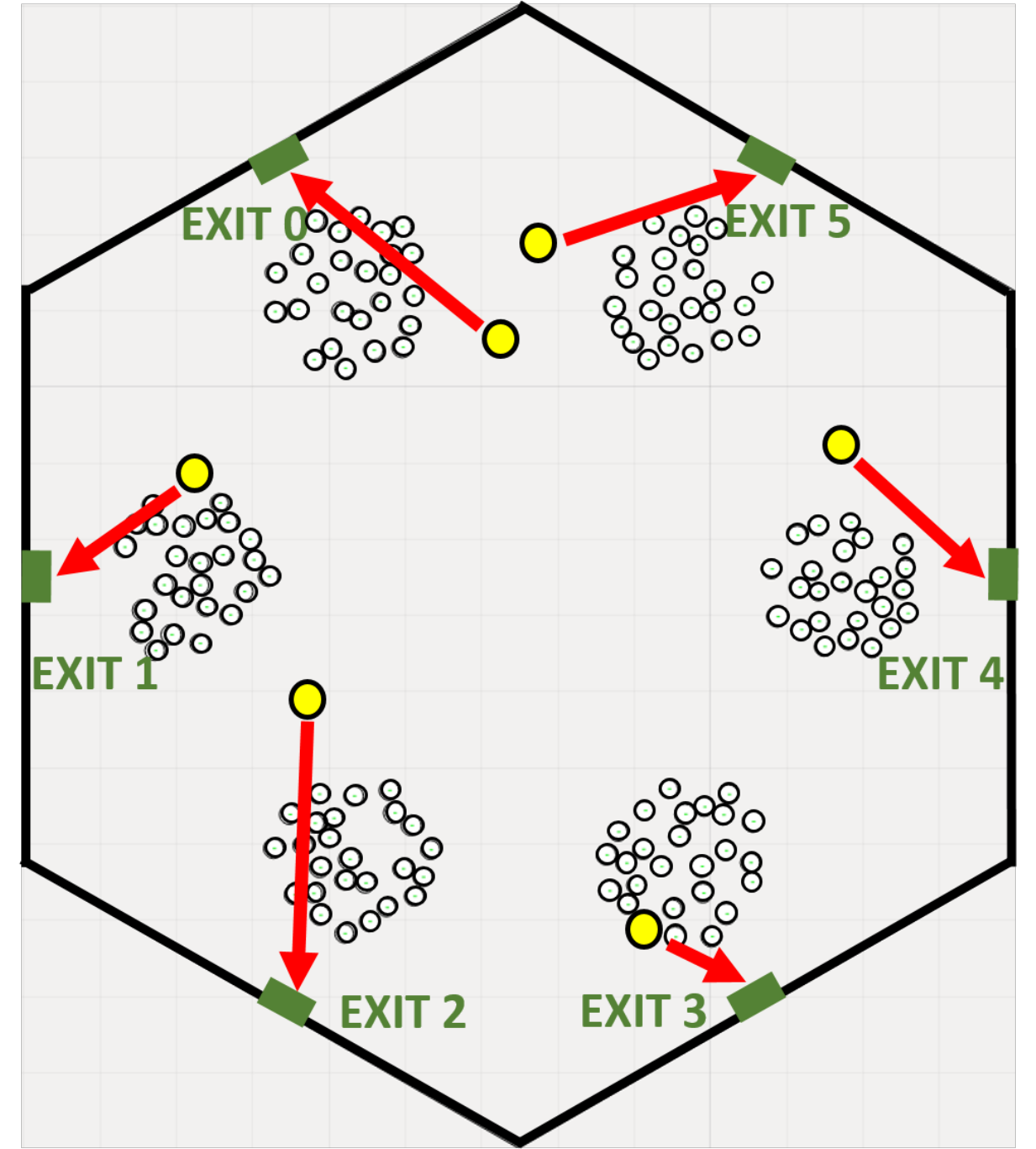}}
\caption[A set of six subfigures.]{The near-optimal evacuation plans with \subref{fig:figure_5a} one guide, \subref{fig:figure_5b} two guides, \subref{fig:figure_5c} three guides,
\subref{fig:figure_5d} four guides, \subref{fig:figure_5e} five guides and \subref{fig:figure_5f} six guides.}
\label{fig:figure_5}
\end{figure}

The resulting near-optimal solutions can be seen in \mbox{Fig.~\ref{fig:figure_5}}. If there is only one guide, it leads half of the exiting agents to the exit on the opposite side of their familiar exit. If there are two guides, the crowd is split into three parts heading towards three different exits. With three guides, the crowd is split into four parts heading towards four different exits. With four guides, the crowd is split into five parts heading towards five different exits. With five guides, we get almost the same solution as with the hidden genes GA, only the starting positions of the guides are slightly altered. With six guides, the sixth guide is put aside to evacuate with the group that is heading to Exit $0$. When adding the extra guide, the GA assigns a starting position and target exit to it so that it does not interfere with the minimum time evacuation plan.

\section{Evacuation of a conference building}\label{sec:conference}

\begin{figure}[htb!]
\centering\includegraphics[width=0.9\textwidth]{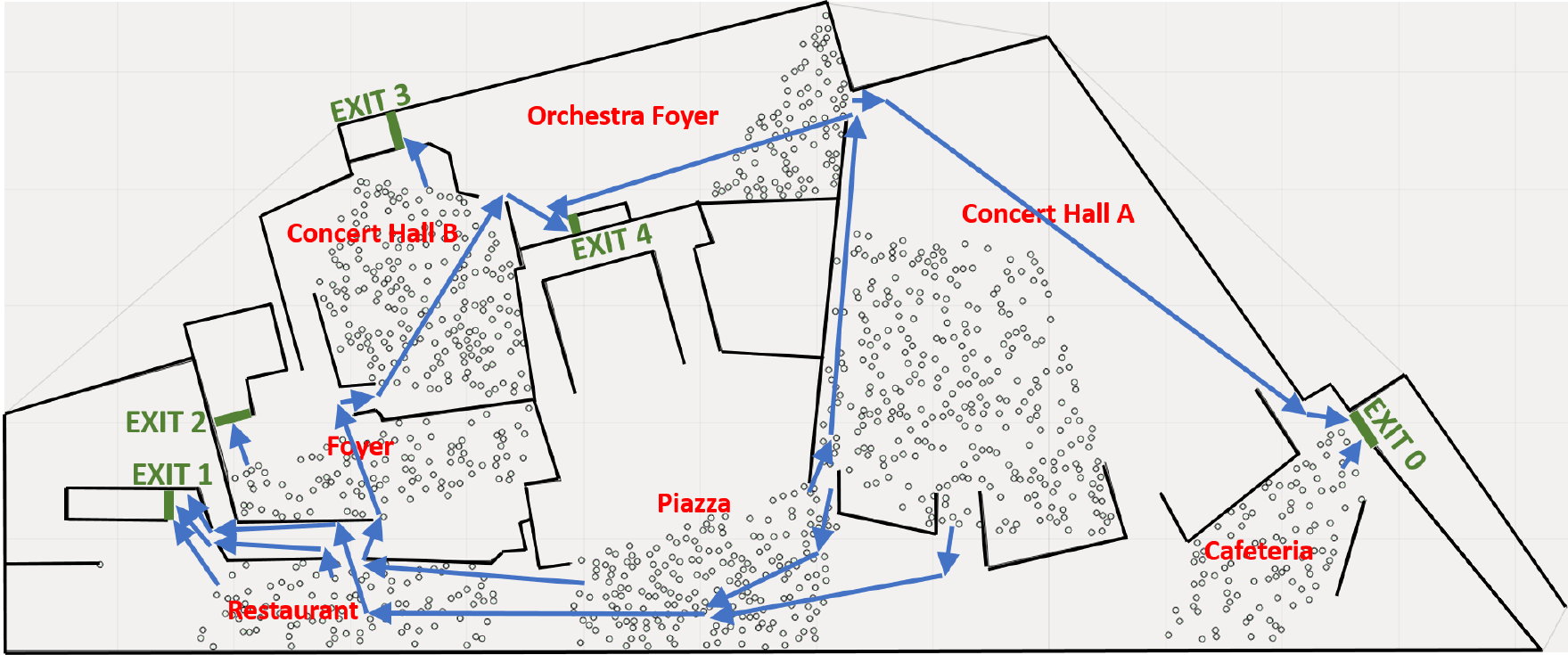}
     \caption{The initial situation of the unguided evacuation of the conference building.}
  \label{fig:figure_6}
\end{figure}

We have now tuned the GA parameters suitable for the test case, and will apply it to the evacuation of a crowd from a conference building in an emergency situation. The initial situation is depicted in Fig.~\ref{fig:figure_6}.

\begin{figure}[htbp!]
\centering
\subfigure[][]{
\label{fig:figure_7a}
\includegraphics[width=0.85\textwidth]{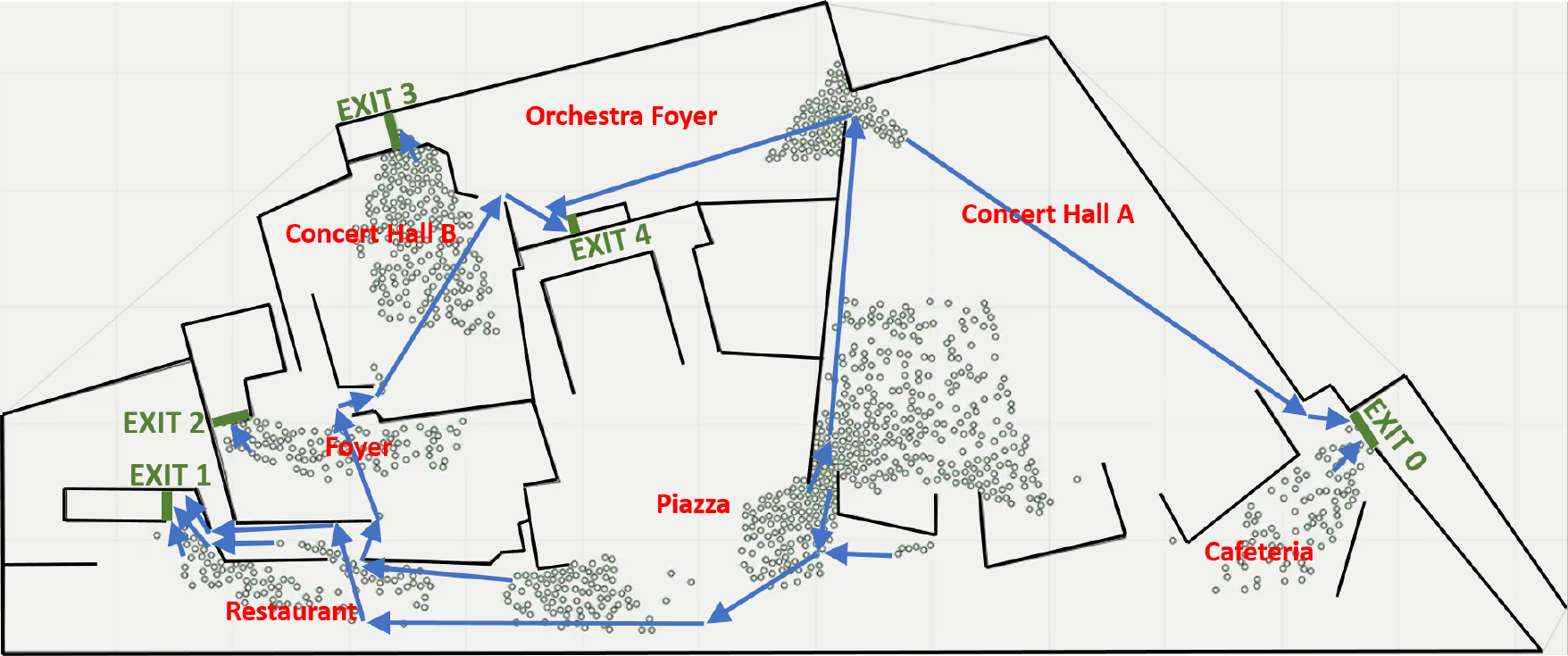}}

\subfigure[][]{
\label{fig:figure_7b}
\includegraphics[width=0.85\textwidth]{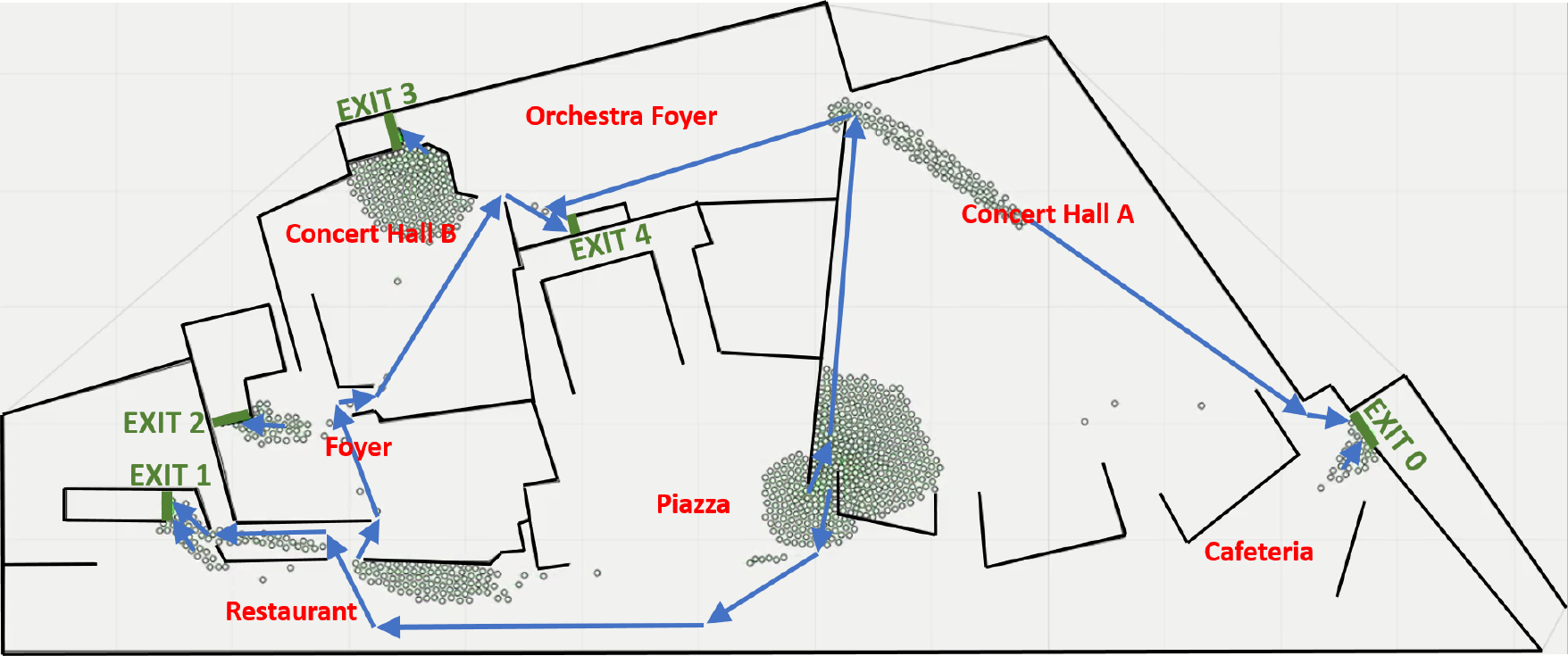}}

\subfigure[][]{
\label{fig:figure_7c}
\includegraphics[width=0.85\textwidth]{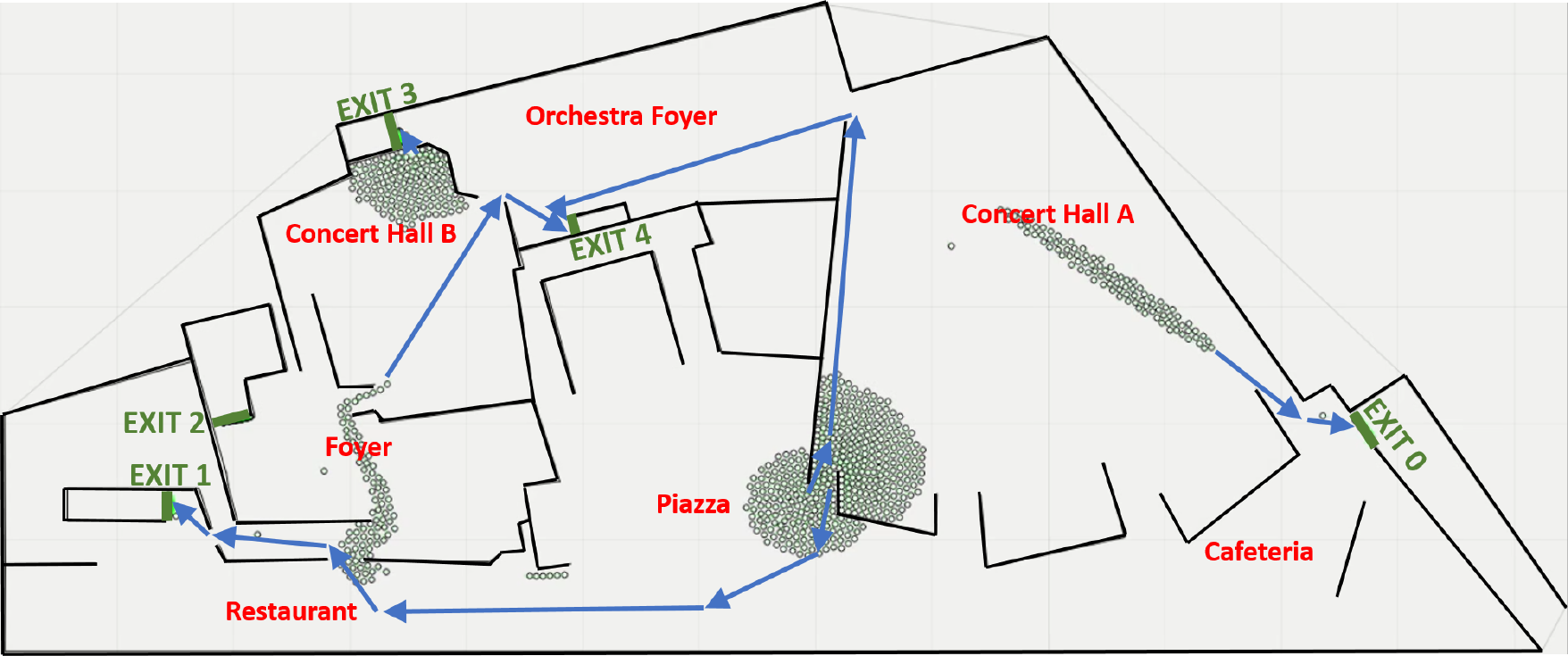}}

\caption[A set of six subfigures.]{Snapshots of one realization of the unguided evacuation of the conference building at \subref{fig:figure_7a} $t=5$ s; \subref{fig:figure_7b} $t=15$ s; \subref{fig:figure_7c} $t=30$ s.}
\label{fig:figure_7}
\end{figure}

\begin{figure}[htb!]
\centering
\subfigure[][]{
\label{fig:figure_8a}
\includegraphics[width=0.85\textwidth]{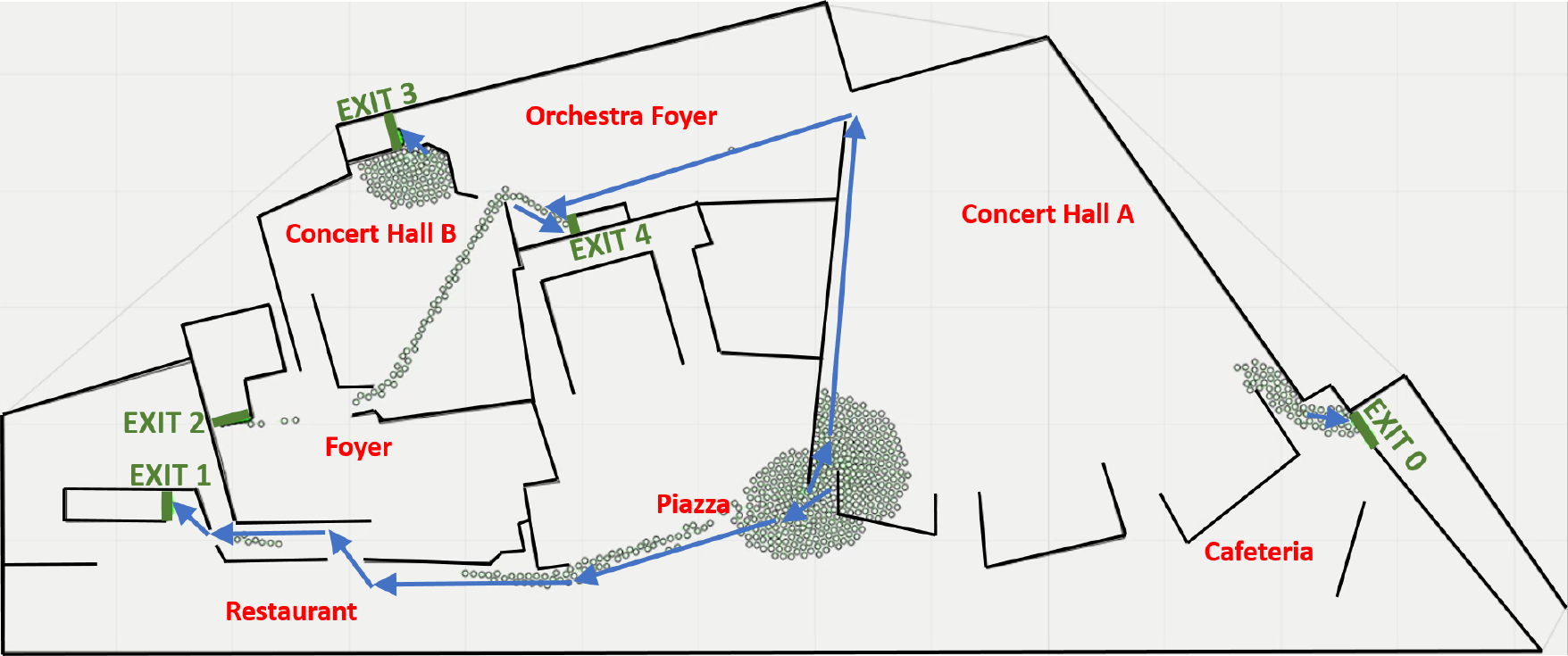}}

\subfigure[][]{
\label{fig:figure_8b}
\includegraphics[width=0.85\textwidth]{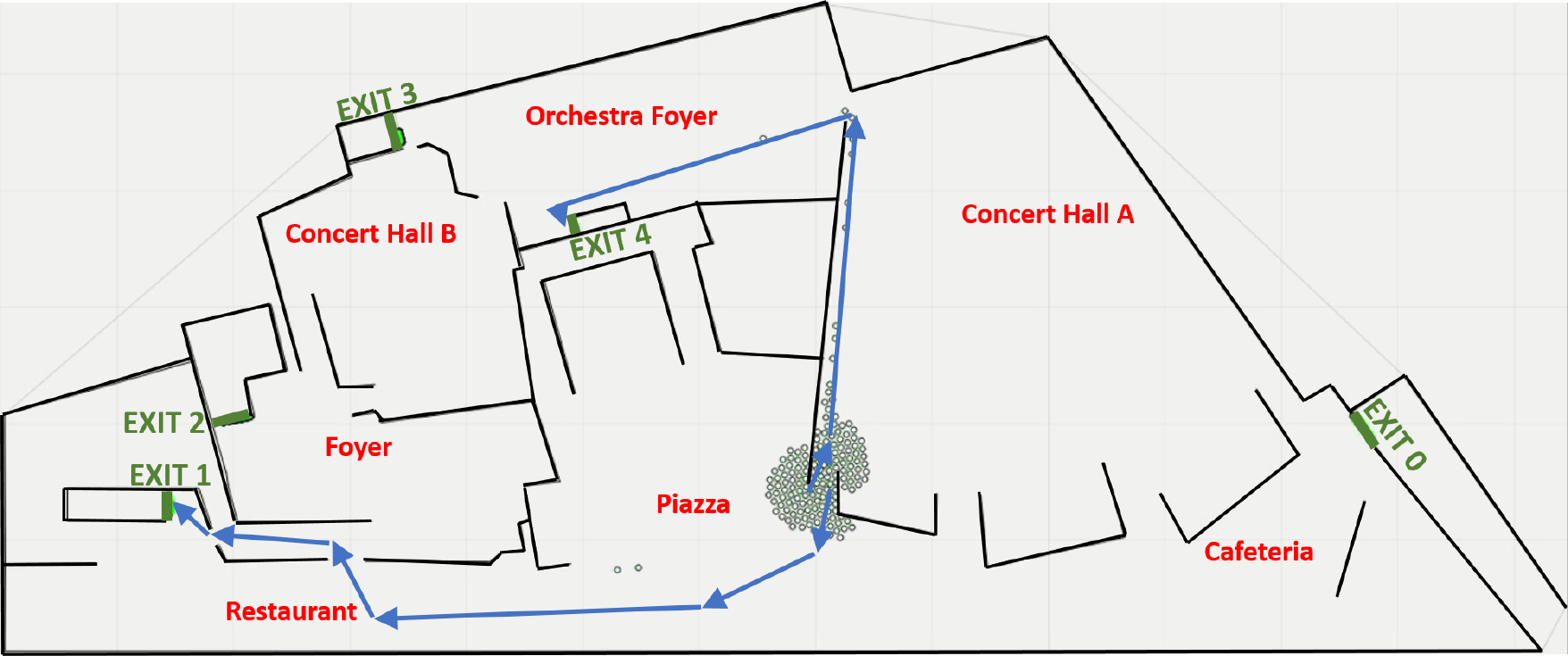}}
\caption[A set of six subfigures.]{Snapshots of one realization of the unguided evacuation of the conference building at \subref{fig:figure_8a} $t=50$ s; \subref{fig:figure_8b} $t=205$ s.}
\label{fig:conference_snapshots_b}
\end{figure}

Initially, the crowd of $1100$ exiting agents is split into subgroups in seven different rooms: Concert Hall A, Cafeteria, Piazza, Restaurant, Foyer, Concert Hall B and Orchestra Foyer. There are five exits in the building. All agents in a subgroup have the same familiar exit towards which they initially intend to head (depicted by the blue arrows in Fig.~\ref{fig:figure_6}). In the evacuation model, the agents move to their target exit using their shortest path; note however that even in a small group the agents may take different (shortest) paths to their target exit. This is the case for the group of agents initially in Piazza, which are heading to Exit 4 in Orchestra Foyer. Half of the group heads first to Concert Hall A and move along its left wall to Orchestra Foyer, while the other half takes the route through Restaurant, Foyer and Concert Hall B. The development of one realization, or scenario, of the unguided evacuation is seen in Figs.\mbox{\ \ref{fig:figure_7a} to \ref{fig:figure_8b}}.

Almost immediately a large jam emerges at the bottom leftmost entrance of Concert Hall A. Half of the exiting agents from Piazza try to enter Concert Hall A, through the same entrance that the exiting agents in Concert Hall A use to enter Piazza. This jam is cleared out very slowly, and it is the main source of inefficiency for the evacuation.

Here, all Assumptions 1-5 are used. We set $r^{guide}=10$ m. It is larger compared to the $5$ m interaction range used for the hexagon-shaped case. We could use the same value in both cases. The parameter seems to have an effect on how close the guide should be positioned to the exiting agents under its influence, and how large these groups of exiting agents are. We here set it larger so that the guide could collect more exiting agents from the large crowd. Perhaps, if it is set smaller here, more guides would be needed to have an effect on the crowd. The exit visibility range is set smaller than the interaction range of the guide, $r^{exit}=9$ m. As a result of this, in some situations, the guides are able to influence exiting agents before they notice a nearby exit. So, the relation between the values of $r^{guide}$ and $r^{exit}$ can have an effect on the exiting agents' exit choice in situations where both a guide and an exit are nearby. Otherwise, the same evacuation model parameter values are used as in the hexagon-shaped case. 

\subsection{Numerical results}

\begin{figure}[ht!]
\centering\includegraphics[width=0.9\textwidth]{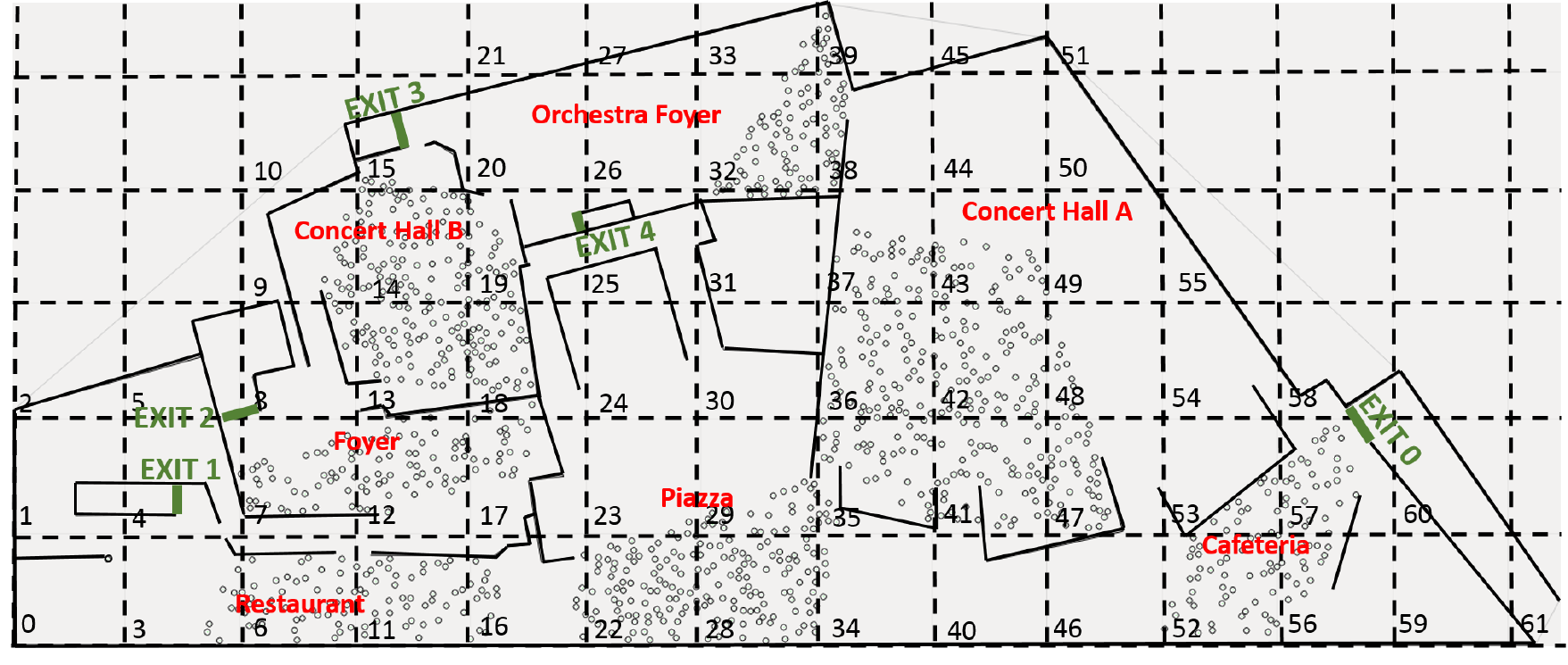}
     \caption{Feasible starting grid cells for the guides in the conference building.}
  \label{fig:figure_9}
\end{figure}

In the GA, we set the number of genes in a chromosome to $10$, meaning that the solution can have a maximum of $10$ guides. When the GA has converged, all solutions in the final population have less than $10$ genes active, which strongly indicates that the optimal solution should have less than $10$ guides. The feasible starting grid cells are obtained again by discretizing the building into sixty-two  $10$ m $\times$ $10$ m grid cells; see \mbox{Fig.~\ref{fig:figure_9}}. The five exits in the conference building are the feasible target exits.

\begin{figure}[ht!]
\centering\includegraphics[width=0.9\textwidth]{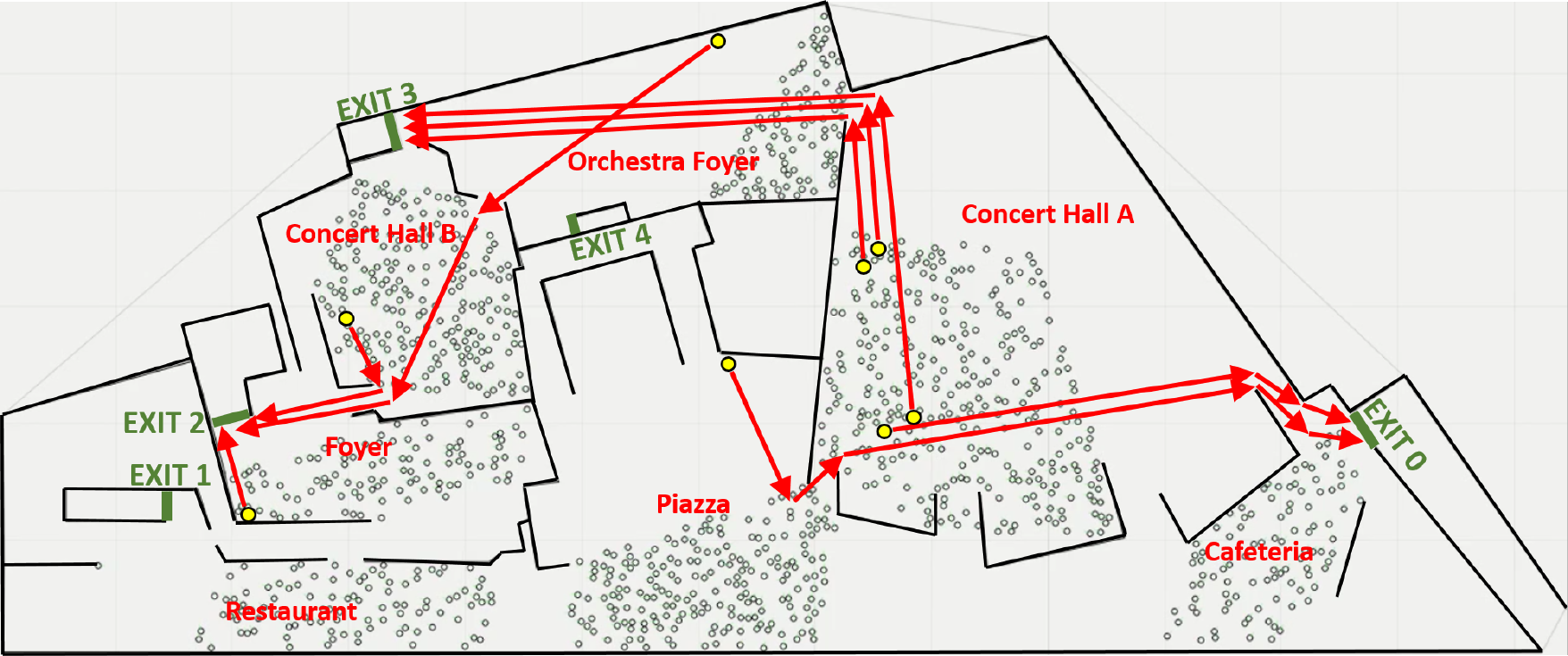}
     \caption{The near-optimal evacuation plan of the conference building.}
  \label{fig:figure_10}
\end{figure}

As in the test case, the crossover probability is set to $0.85$, the mutation probability to $0.10$, and the population size to $40$, with $30$ samples of each chromosome. The two worst individuals are always replaced with the two best individuals of the previous generation. The near-optimal solution given by the hidden genes GA is illustrated in \mbox{Fig.~\ref{fig:figure_10}}.

The total number of active guides is eight (depicted by the yellow circles). They head to their target exits (Assumption 2) along the routes depicted by the red arrows. The exiting agents start to follow their closest guides, within the $r^{guide}$ range, and do not switch to follow another guide (Assumptions 3 and 4). One guide is set in the Orchestra Foyer to head towards Exit 2. Four guides are in Concert Hall A, to herd the large crowd. Out of those four, three are heading towards Exit 3, and one towards Exit 0. In the upper part of Piazza, there is a guide, which leads half of the exiting agents in Piazza to Exit 0. The other half of the exiting agents in Piazza are initially free to move towards Exit 4 using the route through Restaurant, Foyer and Concert Hall B. The exiting agents located in Cafeteria, Restaurant and Foyer move as they would without guidance. There is also one guide in Foyer and another in Concert Hall B that lead half of the exiting agents from Concert Hall B to Exit 2. The other half of the exiting agents in Concert hall B go to Exit 3 without guidance. Finally, there is one guide located in the Orchestra Foyer, and it is heading towards Exit 2.

\begin{figure}[ht!]
\centering\includegraphics[width=0.6\textwidth]{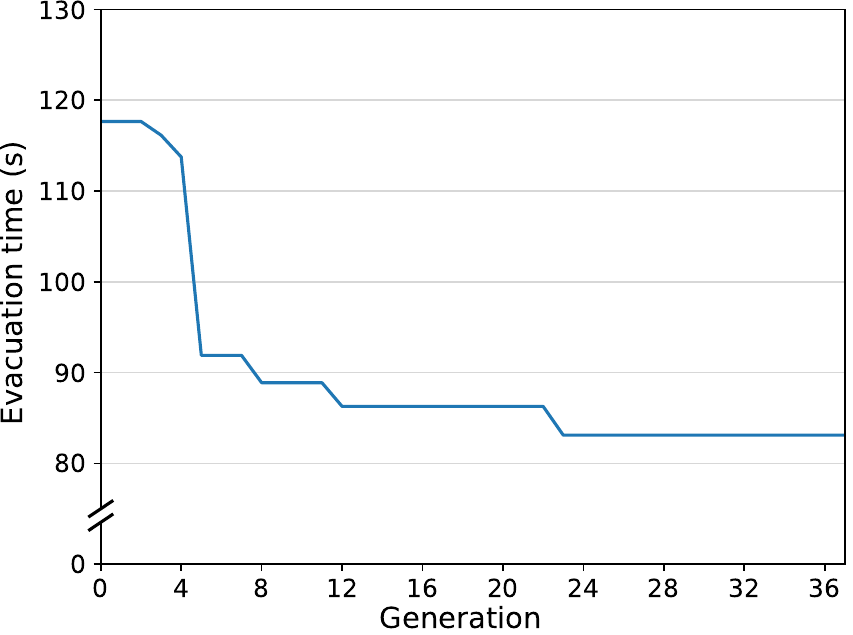}
     \caption{The sample mean evacuation time of the best solution of each generation of the GA, when applied to the conference building case.}
  \label{fig:figure_11}
\end{figure}

The convergence pattern of the GA can be seen in Fig.\mbox{~\ref{fig:figure_11}}. Starting from the $23$rd generation, the best solution does not change for $15$ generations, and the GA is considered to have converged.

\begin{figure}[htbp!]
\centering
\subfigure[][]{
\label{fig:figure_12a}
\includegraphics[width=0.85\textwidth]{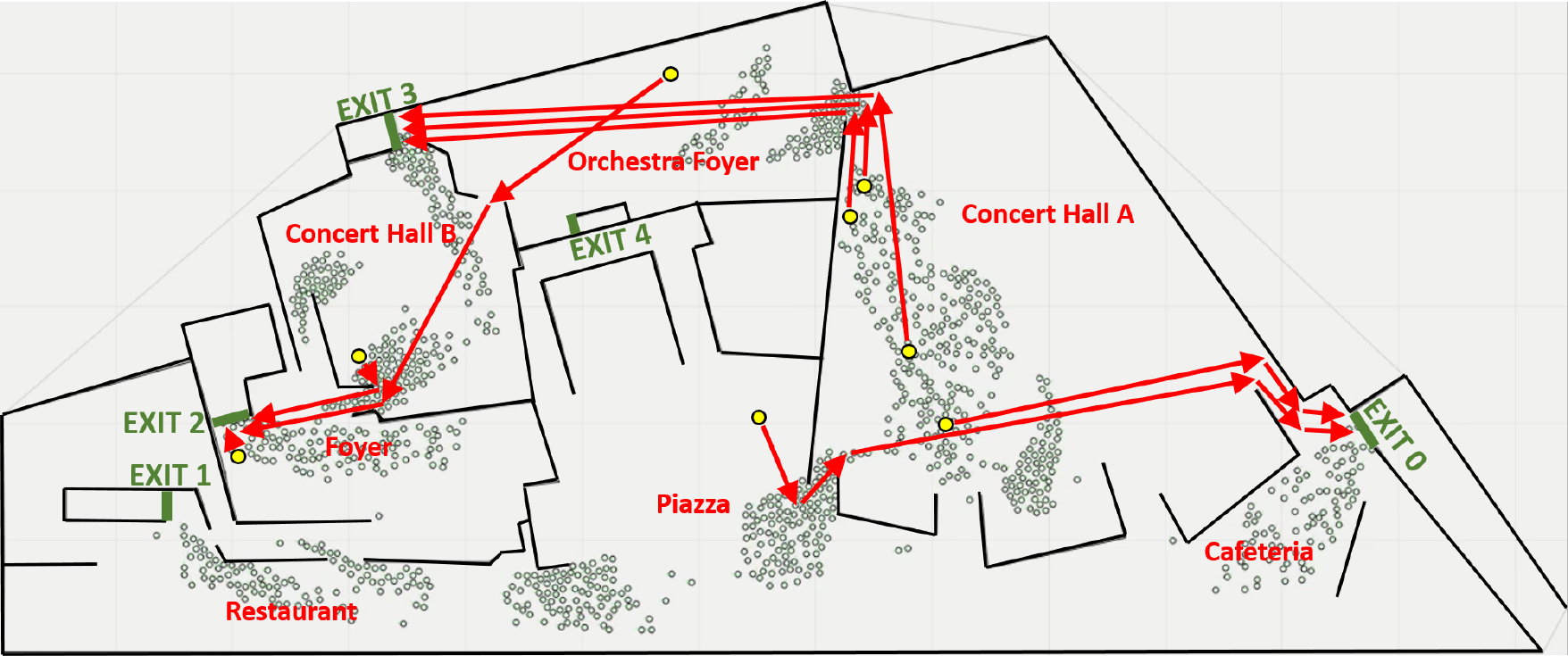}}

\subfigure[][]{
\label{fig:figure_12b}
\includegraphics[width=0.85\textwidth]{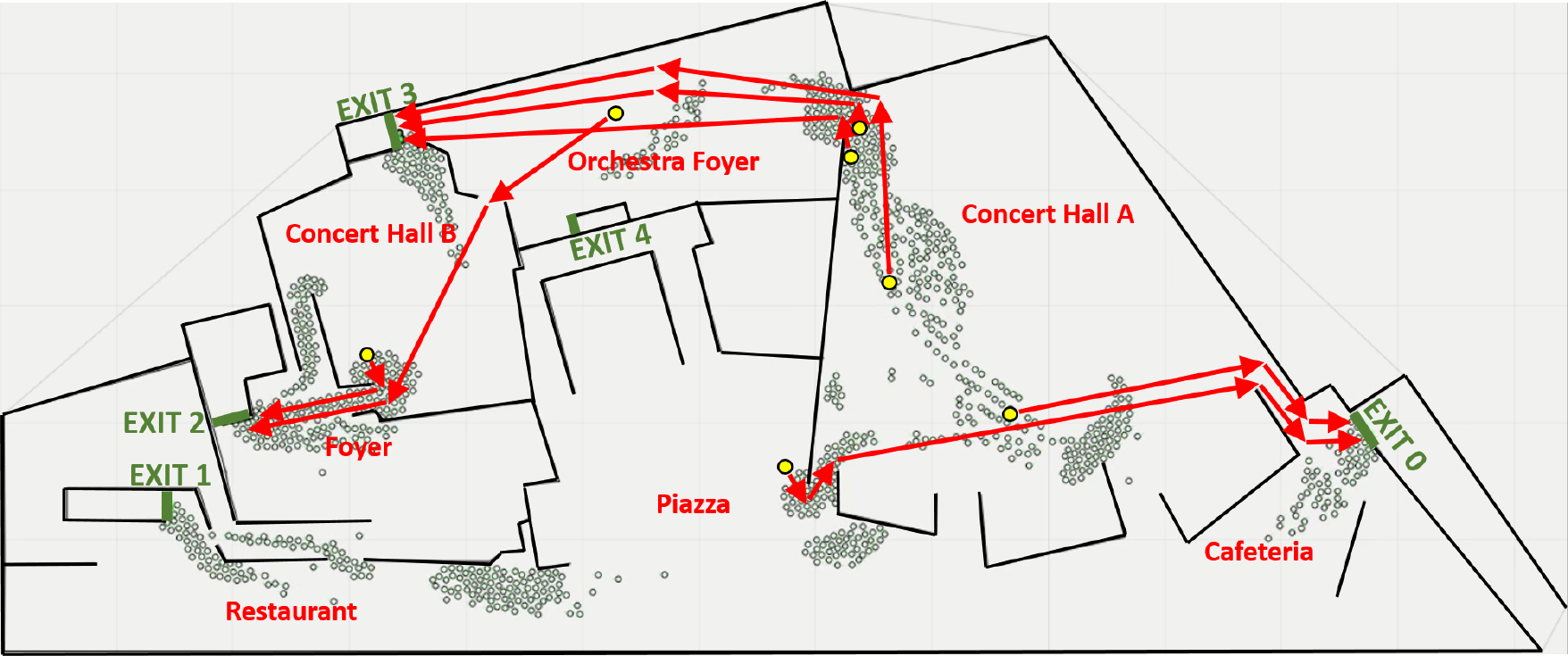}}

\subfigure[][]{
\label{fig:figure_12c}
\includegraphics[width=0.85\textwidth]{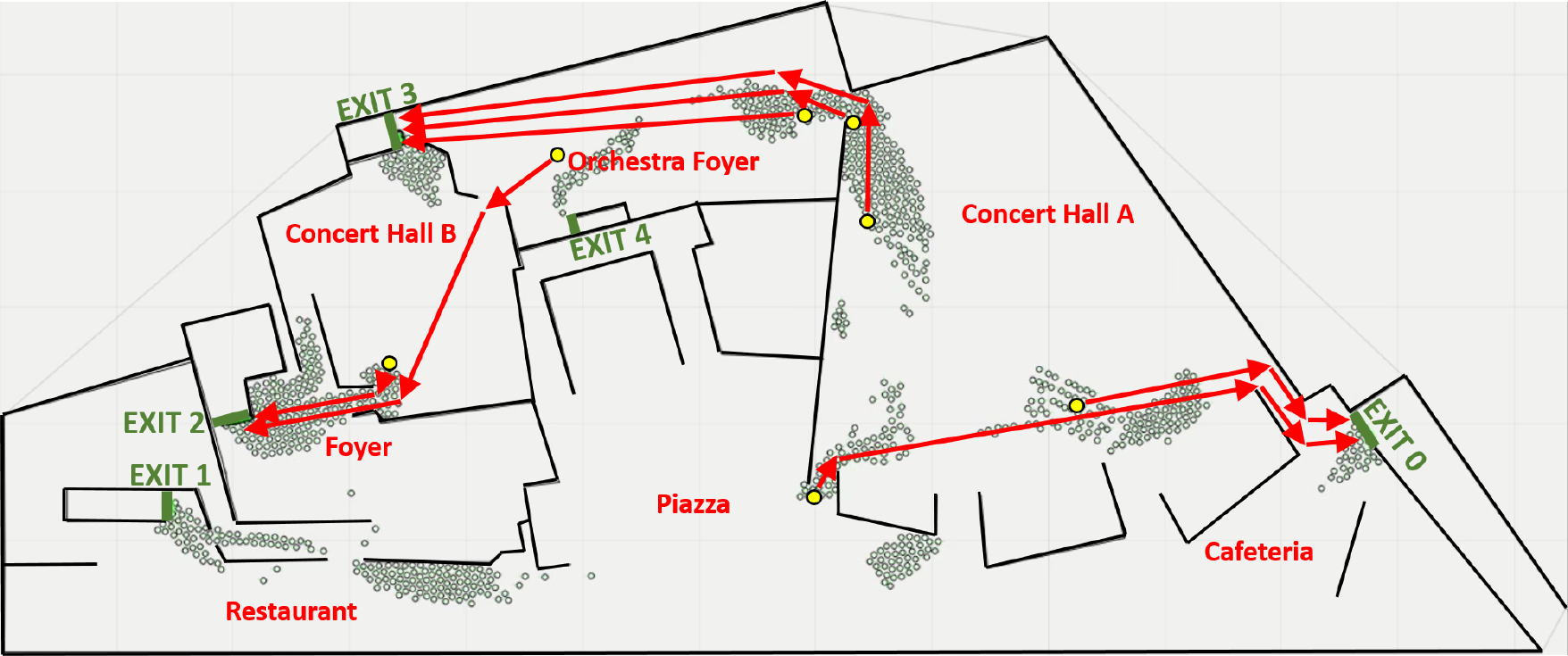}}

\caption[A set of six subfigures.]{Snapshots of one realization of the near-optimal guided evacuation of the conference building at \subref{fig:figure_12a} $t=5$ s; \subref{fig:figure_12b} $t=10$ s; \subref{fig:figure_12c} $t=15$ s.}
\label{fig:figure_12}
\end{figure}

\begin{figure}[ht!]
\centering
\subfigure[][]{
\label{fig:figure_13a}
\includegraphics[width=0.85\textwidth]{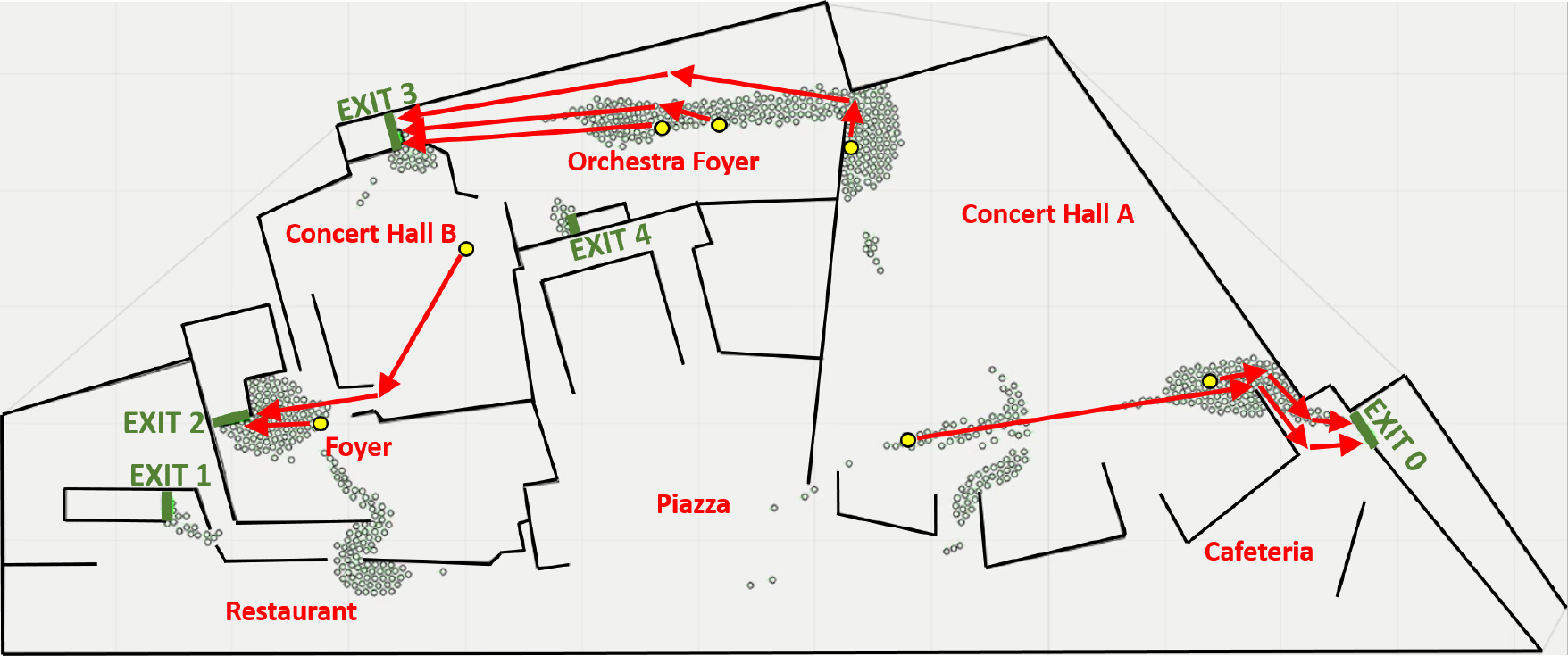}}

\subfigure[][]{
\label{fig:figure_13b}
\includegraphics[width=0.85\textwidth]{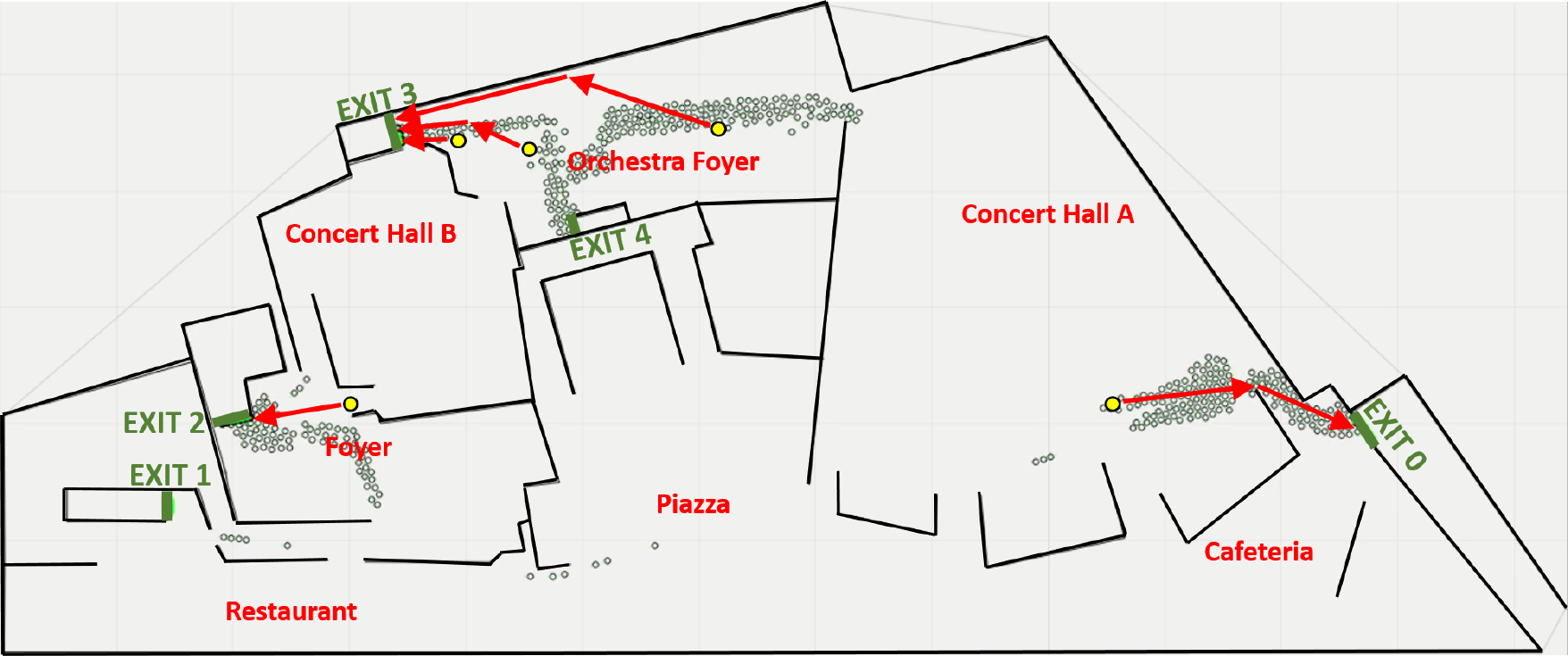}}
\caption[A set of six subfigures.]{Snapshots of one realization of the near-optimal guided evacuation of the conference building at \subref{fig:figure_13a} $t=25$ s; \subref{fig:figure_13b} $t=40$ s.}
\label{fig:figure_13}
\end{figure}

In the unguided situation, a large jam forms at the bottom leftmost entrance to Concert Hall A. It is evident that it should be solved to increase efficiency. However, the question is where to redirect this part of the crowd, as not to create a jam somewhere else. In Figs.\ \ref{fig:figure_12a} to \ref{fig:figure_13b} the full evolution of one realization of the near-optimal guided evacuation is illustrated with snapshots.

There are many changes to the unguided evacuation. First and foremost, a big jam never emerges at the bottom leftmost entrance of Concert Hall A. This is because the crowd in Concert Hall A is taken to Exit 3 through Orchestra Foyer, and the rest to Exit 0 via the rightmost door of Concert Hall A. Only a couple of exiting agents from Concert Hall A are left out without a guide and move by themselves to their familiar exit, Exit 1.

A large part of the crowd is heading towards Exit 3, and no guide is going towards Exit 4. However, guiding exiting agents to Exit 4 would be unnecessary, as about half of the exiting agents initially heading to Exit 3, once they are in the Orchestra Foyer, detect Exit 4 by themselves, and head there instead of Exit 3 (Assumption 5).

Now without guidance, the exiting agents from Piazza would be heading to Exit 4, which would cause too much traffic there. However, in the near-optimal solution, a guide takes half of the exiting agents from Piazza to Exit 0 through Concert Hall A. The second half of the exiting agents in Piazza that head to Exit 4 through Restaurant, Foyer and Concert Hall B are stopped at the Foyer and redirected to Exit 2.

In a more complicated model, the guide could be set to wait in place for exiting agents crossing paths with it later on. Here, the optimization model circumvents this lack of feature, by setting guides to start moving from farther away, to cross paths with exiting agents just at the right time. This is illustrated by the guide from the upper right corner of Orchestra Foyer, that crosses paths with the exiting agents from Piazza at Foyer, and redirects them to Exit 2. To conclude, the main feature of a good guidance seems to be that all exits are utilized, while none of them are overutilized. Also, jams or counterflows should not either emerge at other parts of the building.

\begin{figure}[ht!]
\centering\includegraphics[width=0.6\textwidth]{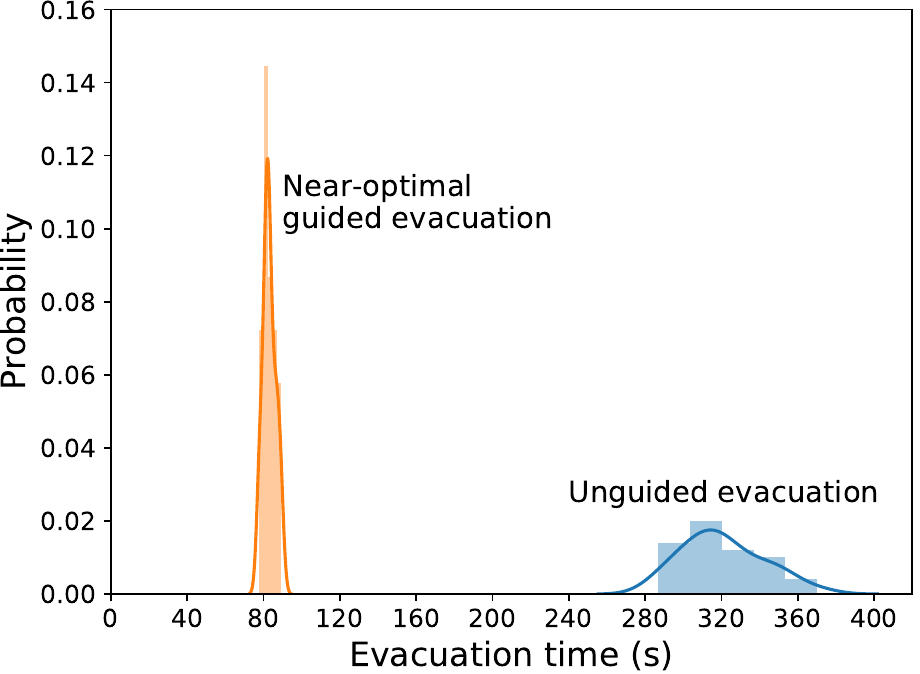}
     \caption{Histograms (staple diagrams) and kernel density estimates (solid lines) of the evacuation times of 30 scenarios.}
  \label{fig:figure_14}
\end{figure}

Fig.\ \ref{fig:figure_14} shows a histogram and a kernel density estimate of the evacuation time of both the unguided and near-optimal evacuation. For the unguided evacuation, the sample mean is $320.72$ s, whereas the sample standard deviation is $20.75$ s. For the near-optimal evacuation, the respective values are $83.11$ s and $3.23$ s. The improvement is tremendous, as the quantities are only about $25$ \% and $15$ \%, respectively, of those values of the unguided evacuation. The improvement in standard deviation probably is due to the near-optimal guidance solving the large jam witnessed in the unguided evacuation. The nonlinear physical forces come into play in close contact of agents. When a large crowd is jammed at a bottleneck, the agents in it are in close contact for a prolonged time. Hence, even small movement fluctuations can affect the evacuation time. This has also been experimentally verified in \mbox{\citep{latticegassimulation, staticfloorfield}}. In Appendix B, we have further analyzed the effect of stochasticity on the problem by using metrics from the literature of stochastic programming \mbox{\citep{birge1982}}.

Also, from another point of view, even though the initial positions of the agents are fixed for all simulations, as time goes on, due to the random force term, the positions will differ between scenarios. Thus, in some sense, the near-optimal evacuation plan is robust to reasonable variations in agents' positions.

\section{Behavioral considerations}\label{sec:behavioral}

To emphasize, what we have in mind with our model, or the basic question we ask is this: what will happen in a complex building containing a large crowd, after the sound of a whistle or a fire alarm in a serious situation, when there is enough trained yellow-coated rescue personnel, whose members have enough skill and authority to guide the crowd out of the building as soon as possible. So, our main modeling question here is how to coordinate the guides so that they manage with their task in an optimal way.

A typical concern in modeling crowd evacuation, is that the crude assumption of focusing only on physical movement may potentially lead to an underestimation of the time for the crowd to reach safety, since evacuees are likely to engage in a variety of activities during the operation \citep{gwynne2016}. We believe that in our situation guided by trained personnel, such evacuation activities, although can be important, are very rare, since people in serious situations are willing to follow authorities \citep{proulx2002, nishinari}.

Even more to the point, it does not matter if some individuals, at some time, would have turned to the right, or stopped for a while, since now guides more or less control their movement. In fact, even for a large group of people, a few guides are enough to control the whole crowd to their target \citep{humangroup,multigridmodel}. For a more irregular building geometry, where the crowd is more scattered, more guides are needed. In that case, the coordination between guides becomes very important to facilitate the evacuation. A useful quantitative information that a model should deliver in this case is a suitable macroscopic one, such as the evacuation time of the crowd. It is in fact less prone to globally unnecessary details and fluctuations than microscopic information \citep{helbing2013}. Our paper solves the number of guides, their initial positions and exit assignments to minimize the crowd evacuation time in a single although complicated optimization problem.

Nevertheless, we acknowledge the importance of taking into account behavioral aspects in crowd evacuation modeling. It has been the focus of our previous research \citep{safetyscience, ownphysreve, ownca, ownphysica}. For example, in \citep{ownphysica} we model crowd behavior in an exit congestion. We couple a local decision-making model to the social force model. The decision-making model is based on behavioral assumptions that are verified in our experiment with real humans \citep{safetyscience}. With our integrated treatment of behavioral and physical aspects, we are able to simulate when, why and how typical phenomena of an evacuation through a bottleneck occur. Most importantly, we attain non-monotonous speed and kinetic pressure patterns near exits in threatening situations. These kind of behavioral phenomena are interesting as such. However, to include them into the optimization would require us to increase the already very heavy computation.

Instead of studying a single agent deviating from its usual movement, it would be interesting to study the possibility of a larger group doing so. For example, it could happen like this: an agent decides to go to another exit and a large 20\% part of the crowd decides to follow this agent. However, to add this behavior to a crowd movement model takes careful calibration. It is not a simple task to alter the microscopic dynamics to generate this effect. Moreover, a larger amount of Monte Carlo simulations would probably be needed to estimate the mean of the evacuation time for a given evacuation plan. Perhaps, an easier way than for the model to generate this effect, is to construct representative scenarios. They would describe all drastically different alternative ways the evacuation could develop. In one of the scenarios, 20\% of the crowd would go to another exit as opposed to where it goes in the other scenarios. Then, the optimal evacuation plan would be calculated over all these scenarios.

\section{Implementation details and performance}\label{sec:implementation}

Let us first start by discussing the implementation details of the evacuation model. In it, the shortest paths that agents use to move to the exits were calculated using the detailed method presented in depth in \mbox{\citep{kretz}}. The method involves solving the continuous shortest path to an exit. It is solved using the fast marching method \mbox{\citep{fastmarchingmethod}}, which first discretizes the continuous space into a meshgrid. Then the method works almost like Dijkstra's algorithm for finding shortest paths between nodes in a graph. The solution is a distance map from each point, or grid, in the building to the exit. The direction to which an agent should head towards at each point can be calculated using the gradient direction. The distance maps, i.e., the distances from each grid to every exit are calculated and stored before the optimization simulations.

Interaction forces in the social force equations should be calculated between all agents, but to speed up the computation, we subdivided the simulation domain with a blocklist algorithm with a cut-off distance of $3$ m \mbox{\citep{blocklist}}. This should not affect the crowd dynamics, since the social force term is exponentially decaying.

When using numerical simulations to calculate the sample mean evacuation time, we use a pseudorandom number generator to generate realizations of the random force terms. For each scenario, we store the seed of the pseudorandom number generator. Thus, the realizations of the random force terms are replicable. Then, for a specific seed, the social force equations are solved with the Velocity Verlet numerical integration scheme with the time step $\triangle t = 0.01$ s; see, e.g., appendix of \mbox{\citep{ownphysica} for more details}.

The evacuation model is implemented in Python code. Some of the core parts of the code are written as Numba-decorated functions, which translates Python functions to optimized machine code at runtime. Numba-compiled numerical algorithms in Python can approach the speeds of C or FORTRAN \mbox{\citep{numba}}. Whereas, the GA is implemented in Bash script that calls the crowd simulation scripts written in Python. For reproducibility, all codes are published \mbox{\citep{articlecode}}.

We ran the numerical experiments on the Aalto University high-performance computing cluster Triton. A single generation of the GA was run in parallel on Triton using its computing nodes that are Intel Xeon X5650 $2.67$ GHz with $48$ GB or $96$ GB memory, and Xeon E5 2680 v2 $2.80$ GHz with $64$ GB or $256$ GB memory. One generation in the GA requires $1200$ simulations, because a generation contains $40$ solutions, each which have to be run for all $30$ scenarios. Due to the user quotas set for Triton users, we only ran $300$ simulations in parallel, which means that one generation had to be run in four parts. A single simulation of the evacuation of the hexagon-shaped space could take up to $5$ min, while one simulation of the evacuation of the conference could take up to $1$ h. Thus, the simulation of one generation took about $20$ min and $4$ h, respectively. We ran the GA for $26$ and $38$ generations, for the test case and the conference building case, respectively. Thus, for the test case, the GA converged to the near-optimal solution in $8$ h $40$ min, whereas, for the conference building case it converged in $152$ h, or $6$ days $8$ h.

\section{Conclusion}\label{sec: conclusion}

We have studied the problem of minimizing the evacuation time of a crowd from a complex building using rescue guides. The problem is first formulated as a stochastic control problem, where the objective is the expected evacuation time, and the optimization variables are the number of guides and their routes defined as origin-destination pairs. The system equations are the equations of motion, given by the social force model, and the rules of interaction for exiting agents and guides. The problem is then reformulated as a scenario optimization problem, which we solve with a hybrid numerical simulation and GA approach. With it, we are able to solve the number of guides, their initial positions and exit assignments to minimize the crowd evacuation time in a single optimization.

Typically, to solve an optimization problem with a GA, the algorithm parameters are tuned manually in a problem-specific manner. In our study, we do this by constructing a test case, the hexagon-shaped space in \mbox{Sec.~\ref{subsec:hexagon}}, for which we know beforehand what the expected minimum evacuation time should be. Then we extensively try different parameter configurations to get the GA to converge as efficiently and accurately as possible to the near-optimal solution. This GA parameter configuration is also used for solving the near-optimal evacuation plan for the conference building.

Moreover, it is known that using GAs for problems with large solution space and large number of local minima, introducing noise to the fitness function, and evaluating it by taking multiple samples, improves convergence to the global minimum \mbox{\citep{noisyfunction}}. Our problem is inherently noisy, and we perform multiple Monte Carlo simulations to calculate the expected evacuation time for a given evacuation plan. Thus, this may assure that our GA does not get stuck in a local optimum.

In both cases studied in this paper, the improvement with the near-optimal plan is dramatic, the sample mean evacuation time is only about $25$ \% of that of the unguided evacuation. Moreover, for the conference building case, we also compared standard deviations, and it was only about $15$ \% of that of the unguided evacuation. This is probably due to the near-optimal evacuation plan solving all major jams, which decreases the nonlinear physical effects, and hence small deviations in agents' movement does not affect the evacuation time so much.

It is interesting to see how the near-optimal solutions take into account physical phenomena like counterflows, jam formation and flow at bottlenecks being a nonlinear function of crowd size. There are computationally faster evacuation models, like cellular automata and network flow models that are good for planning large-scale evacuations \mbox{\citep{lovosflow,abdelghany}}. However, they cannot model the physics of crowds evacuating from complex confined spaces.

However, it should also be noted that the optimization procedure is computationally very slow. So, as such, this procedure could not be used for online optimization. Faster computation could be achieved by using an implicit integration scheme, where the step size can be set in some cases even 40 times larger than with an explicit integration scheme \citep{implicitcrowd}. Alternatively, a position-based dynamics approach could be used, which has been shown to produce real-time simulations \citep{positionbased}. Another interesting avenue for future research would be to use neural networks to deal with the large state space \mbox{\citep{neurodynamic}}.

In future research, other objective functions could be used as well, and studied how the optimal evacuation plans differ from that of ours. An evacuation plan should be both fast and safe \citep{lovos}. If we want to simultaneously take these two objectives into account, we can solve the problem with multi-objective optimization \citep{moea}. As noted before, one form of risk are rare events that dramatically slow down the evacuation. To account for this, we cannot only minimize the mean evacuation time, but we also have to minimize the variance or some other risk-related measure related to the evacuation time distribution. We could also include physical risks like pressures in the crowd or avoidance of dangerous areas in a building. On the other hand, if we are considering an unhurried evacuation, objectives related to the quality of service can be used, e.g., minimization of the average evacuation time or distance travelled by the evacuees, or time spent in congestions.

\section*{Acknowledgements}

This study was first funded by a grant from the Foundation for Aalto University Science and Technology, and later with a grant from the Finnish Science Foundation for Technology and Economics. The calculations in this study were performed using computer resources within the Aalto University School of Science "Science-IT" project. We wish to thank our summer assistant Jaan Tollander de Balsch, who considerably helped us develop the simulation codes.

\section*{Declaration of interests}
The authors declare that they have no known competing financial interests or personal relationships that could have appeared to influence the work reported in this paper.

\appendix

\section{Social force model parameters}\label{appendix: appendixa}

The social force model parameters have been validated against data, and the collective phenomena observed with the model are robust to reasonable parameter variations \citep{powerlaw, fdsevac}. In our study, the initial positions $\mathbf{x}_i^0$, radii $r_i$, masses $m_i$, and desired speeds $v_i^{des}$, for exiting agents $i \in N$, are fixed for all simulations. Before fixing them, the parameters $m_i$, $r_i$, and $v_i^{des}$ are drawn from a truncated normal distribution with a cutoff at three times the standard deviation. The mean and standard deviation are $73.5$ kg and $8.0$ kg, $0.255$ m and $0.035$ m, and $1.25$ m/s and $0.3$ m/s, respectively for $m_i$, $r_i$ and $v_i^{des}$. Whereas, for guide agents $g \in G$, we set $m_g=80$ kg, $r_g=0.27$ m, and $v_g^{des}=1.15$ m/s. The reaction time is $\tau^{reac}=0.5$ s.

Next, we describe the terms of the social force model. Our derivation of the social force term $\mathbf{f}_{ij}^{soc}, i,j \in N \cup G, i \neq j$, follows closely that of \citep{powerlaw}. There, it was inferred from a large public data set that the interaction energy associated with the repulsive social force follows a power law with a sharp truncation at large $\tau$,
\begin{equation}\tag{A.1}
E(\tau) = \dfrac{k_{soc}}{{\tau}^2} e^{-\tau / {\tau}_0}.
\label{eq:interactionenergy}
\end{equation} 

Here, $k_{soc}$ is a constant that sets the units of energy. It is not that clearly documented, but by examining the codes provided in the supplemental material of \citep{powerlaw}, we deduce its value to be $k_{soc}=1.5m_i$. The parameter $\tau_0$ is the interaction time horizon, and it is set to $3$ s. The collision time of two agents $i$ and $j$ is denoted with $\tau$. So, Eq.~\eqref{eq:interactionenergy} defines the interaction energy of a pair of agents, which are on a collision course. This energy is directly related to the social force $\mathbf{f}_{ij}^{soc}$ experienced by agent $i$ due to the interaction with another pedestrian $j$. In particular,
\begin{equation}\tag{A.2}
\mathbf{f}_{ij}^{soc} = - {\nabla}_{{\mathbf{x}}_{ij}} E(\tau).
\label{eq:interactionforce}
\end{equation}

At any given simulation step, we estimate $\tau$ by linearly extrapolating the trajectories of the pedestrians $i$ and $j$ based on their current velocities, $\mathbf{v}_i$ and $\mathbf{v}_j$, and position vectors, $\mathbf{x}_i$ and $\mathbf{x}_j$. Specifically, a collision is said to occur at some time $\tau$, if the corresponding circles of the pedestrians of radii $r_i$ and $r_j$ intersect. If no such time exists, the interaction force $\mathbf{f}_{ij}^{soc}$ is a zero vector $\bm{0}$. Otherwise, $\tau = \dfrac{b-\sqrt{d}}{a}$, where $a={\lVert \mathbf{v}_{ij} \rVert}^2$, $b=\mathbf{x}_{ij} \cdot \mathbf{v}_{ij}$, $c={\lVert \mathbf{x}_{ij} \rVert}^2 - (r_i+r_j)^2$, and $d=b^2-ac$. Here, $\mathbf{x}_{ij}=\mathbf{x}_i-\mathbf{x}_j$ is their relative displacement, and $\mathbf{v}_{ij}=\mathbf{v}_i-\mathbf{v}_j$ is their relative velocity. By substituting $\tau$ into Eq.~\eqref{eq:interactionforce}, the interaction force can be written explicitly as:
\begin{equation}\tag{A.3}
\mathbf{f}_{ij}^{soc} = - \left[ \frac{k_{soc}e^{\tau/ {\tau}_0 }}{{\lVert \mathbf{v}_{ij} \rVert}^2 {\tau}^2} \left( \frac{2}{\tau} + \frac{1}{{\tau}_0} \right) \right] \left[ \mathbf{v}_{ij} - \frac{{\lVert \mathbf{v}_{ij} \rVert}^2 \mathbf{x}_{ij} - (\mathbf{x}_{ij} \cdot \mathbf{v}_{ij})\mathbf{v}_{ij}}{\sqrt{(\mathbf{x}_{ij} \cdot \mathbf{v}_{ij})^2-{\lVert \mathbf{v}_{ij} \rVert}^2 ({\lVert \mathbf{x}_{ij} \rVert}^2 - (r_{ij})^2})} \right].
\label{eq:explicitinteractionforce}
\end{equation}

\noindent
Here, $r_{ij}=r_i+r_j$. Furthermore, the force $\mathbf{f}_{ij}^{soc}$ is truncated from above to $2000$ N.

The original social force model \citep{socialforce95} includes also a repulsive social force between agents and walls, $\mathbf{f}_{iw}^{soc}$. Because, even though the desired velocity $\mathbf{v}_i^0$ does not point inside walls by construction, the actual velocity $\mathbf{v}_i$ could do so, i.e., an agent could be pushed inside a wall by other agents in the crowd. To avoid this, we use the approach from \citep{obstacles}. In it, the desired velocity $\mathbf{v}_i^0$ is constructed to heavily point away from a wall, when the agent gets close to it.

Physical contact forces come into play, when agents $i$ and $j$ touch each other, $r_{ij}-\lVert \mathbf{x}_{ij} \rVert \geq 0$:
\begin{equation}\tag{A.4}
\mathbf{f}_{ij}^c = k(r_{ij}-\lVert \mathbf{x}_{ij} \rVert) \mathbf{n}_{ij} + c_d \triangle v_{ji}^n \mathbf{n}_{ij} + \kappa (r_{ij} - \lVert \mathbf{x}_{ij} \rVert) \triangle v^t_{ji} \mathbf{t}_{ij},
\label{eq:contactforce}
\end{equation}

\noindent
where $\mathbf{n}_{ij}=(n_{ij}^1,n_{ij}^2)=\mathbf{x}_{ij}/\lVert \mathbf{x}_{ij} \rVert$ is the normalized vector pointing from agent $j$ to $i$, $\mathbf{t}_{ij}=(-n_{ij}^2,n_{ij}^1)$ is the tangential direction, $\triangle v_{ji}^n=-\mathbf{v}_{ij} \cdot \mathbf{n}_{ij}$ is the normal velocity difference, and $\triangle v_{ji}^t = -\mathbf{v}_{ij} \cdot \mathbf{t}_{ij}$ is the tangential velocity difference. The parameters $k=1.2\cdot10^5$ kg/s, $c_d=500$ kg/s, and $\kappa=4.4\cdot 10^4$ kg/(m$\cdot$s) are force constants. The first term in Eq.~\eqref{eq:contactforce} represents a 'body force' counteracting body compression, the second term a 'damping force, which reflects the fact that the collision of two humans is not an elastic one, and the third term is a 'sliding friction force' impeding relative tangential motion. The physical agent-wall interaction force $\mathbf{f}_{iw}^c$ is treated similarly and same parameter values are used.

Finally, the components of the random force vector $\bm{\xi}_i$ follow a truncated normal distribution with mean zero, standard deviation of $0.1m_i$ m/$\text{s}^2$, and are truncated at three times of the standard deviation.

\section{The value of the stochastic solution}\label{sec: appendixb}

Let us evaluate the effect of stochasticity on our optimization problem using terminology from the literature of stochastic programming \citep{birge1982}. First, recall the mathematical definitions from Secs.\ 2 and 3. Then, let us denote $\bm{\xi}_i (t), 1 \leq i \leq n, t \in [0, T_{last}]$, with $\bm{\delta}(t)$. And, because $T_{last}$ depends on ${\omega}_g, {\varepsilon}_g, g \in G$ and $\bm{\delta}$ through the constraint equations, we define: 
\begin{equation}\tag{B.1}\label{eq:phifunction}
\phi({\omega}_g, {\varepsilon}_g, g \in G; \bm{\delta}) := T_{last}.
\end{equation}

The problem of Eq.\ (3) is called the \textit{recourse problem}, and using Eq.~\eqref{eq:phifunction} it can be rewritten:
\begin{align}\tag{B.2}\label{eq:fixedrecourse}
\text{RP} := & \min_{({\omega}_g, {\varepsilon}_g)} \mathbb{E} \big[ \phi({\omega}_g, {\varepsilon}_g, g \in G; \bm{\delta}) \big]; \nonumber \\
& {\omega}_g \subset \bar{\Omega}, {\varepsilon}_g \subset \mathcal{E}, g \in G, \nonumber
\end{align}

\noindent
subject to the constraints defined in Eq.\ (3). We obtained the sample mean $83.11$ s for RP, in the conference building case, using the detailed solution methodology presented in the paper.

Next, we define the deterministic equivalent problem, or the \textit{expected value problem}:
\begin{align}\tag{B.3}\label{eq:evproblem}
\text{EV} := & \min_{({\omega}_g, {\varepsilon}_g)} \phi \big({\omega}_g, {\varepsilon}_g, g \in G; \mathbb{E} \left[ \bm{\delta} \right] \big); \nonumber \\
& {\omega}_g \subset \bar{\Omega}, {\varepsilon}_g \subset \mathcal{E}, g \in G, \nonumber
\end{align}

\noindent
subject to the constraints defined in Eq.\ (3). Recall that the random force terms $\bm{\xi}_i (t), 1 \leq i \leq n, t \in [0, T_{last}]$, are independent and have zero $\bm{0}$ mean. Hence, $\mathbb{E} \left[ \bm{\delta} \right] = \bm{0}$. Furthermore, in the equation of motion Eq.\ (1), the random force appears as an additive linear term. Thus, the term disappears and we are left with the deterministic social force equation. The deterministic equivalent problem is also solved with the solution methodology presented in the paper. The EV solution $({\hat{\omega}}_g, {\hat{\varepsilon}}_g), g \in G$ is seemingly similar to the RP solution. Only the initial position of one of the guides is moved from Concert Hall A to Orchestra Foyer.

The expected evacuation time for solution $({\hat{\omega}}_g, {\hat{\varepsilon}}_g), g \in G$ is:
\begin{equation}\tag{B.4}\label{eq:eev}
\text{EEV} := \mathbb{E} \big[ \phi({\hat{\omega}}_g, {\hat{\varepsilon}}_g, g \in G; \bm{\delta}) \big].
\end{equation}

\noindent
To calculate it, we perform Monte Carlo simulations. We obtain the sample mean of $85.59$ s for EEV. Thus, \textit{the value of the stochastic solution} is:
\begin{equation}\tag{B.5}
\text{VSS} = \text{EEV} - \text{RP} = 83.11 \, s - 85.59 \, s = -2.48 \, \mathrm{s}.
\end{equation}

\noindent 
So, the sample mean of the evacuation time for the EV solution is close to that of the RP solution. Still, this is not the only side to the matter. We experienced difficulties in getting the GA to converge to the optimum of the deterministic equivalent problem. However, because we had solved the stochastic problem first, we knew what we were seeking. Thus, in the end, we made use of the stochastic problem solution, and performed an exhaustive local search to find the deterministic equivalent problem solution. This convergence issue seems to be a more general concern in solving stochastic problems with their deterministic equivalent using a GA \citep{noisyfunction}. For example, it might be that for a deterministic optimization problem having many local optima, but only one global optimum, when adding stochasticity or averaging over many such deterministic problems makes the problem landscape smoother with only a single optimum.

\bibliography{main}

\begin{thebibliography}{48}
\expandafter\ifx\csname natexlab\endcsname\relax\def\natexlab#1{#1}\fi
\providecommand{\url}[1]{\texttt{#1}}
\providecommand{\href}[2]{#2}
\providecommand{\path}[1]{#1}
\providecommand{\DOIprefix}{doi:}
\providecommand{\ArXivprefix}{arXiv:}
\providecommand{\URLprefix}{URL: }
\providecommand{\Pubmedprefix}{pmid:}
\providecommand{\doi}[1]{\href{http://dx.doi.org/#1}{\path{#1}}}
\providecommand{\Pubmed}[1]{\href{pmid:#1}{\path{#1}}}
\providecommand{\bibinfo}[2]{#2}
\ifx\xfnm\relax \def\xfnm[#1]{\unskip,\space#1}\fi
\bibitem[{Abdelghany et~al.(2014)Abdelghany, Abdelghany, Mahmassani \&
  Alhalabi}]{abdelghany}
\bibinfo{author}{Abdelghany, A.}, \bibinfo{author}{Abdelghany, K.},
  \bibinfo{author}{Mahmassani, H.}, \& \bibinfo{author}{Alhalabi, W.}
  (\bibinfo{year}{2014}).
\newblock \bibinfo{title}{Modeling framework for optimal evacuation of
  large-scale crowded pedestrian facilities}.
\newblock {\it \bibinfo{journal}{European Journal of Operational Research}\/},
  {\it \bibinfo{volume}{237}\/}, \bibinfo{pages}{1105--1118}.
  \DOIprefix\doi{https://doi.org/10.1016/j.ejor.2014.02.054}.
\bibitem[{Abdelkhalik(2013)}]{hiddengene}
\bibinfo{author}{Abdelkhalik, O.} (\bibinfo{year}{2013}).
\newblock \bibinfo{title}{Hidden genes genetic optimization for variable-size
  design space problems}.
\newblock {\it \bibinfo{journal}{Journal of Optimization Theory and
  Applications}\/},  {\it \bibinfo{volume}{156}\/}, \bibinfo{pages}{450--468}.
  \DOIprefix\doi{https://doi.org/10.1007/s10957-012-0122-6}.
\bibitem[{Albi et~al.(2016)Albi, Bongini, Cristiani \&
  Kalise}]{invisiblecontrol}
\bibinfo{author}{Albi, G.}, \bibinfo{author}{Bongini, M.},
  \bibinfo{author}{Cristiani, E.}, \& \bibinfo{author}{Kalise, D.}
  (\bibinfo{year}{2016}).
\newblock \bibinfo{title}{Invisible control of self-organizing agents leaving
  unknown environments}.
\newblock {\it \bibinfo{journal}{SIAM Journal on Applied Mathematics}\/},  {\it
  \bibinfo{volume}{76}\/}, \bibinfo{pages}{1683--1710}.
  \DOIprefix\doi{https://doi.org/10.1137/15M1017016}.
\bibitem[{Aub{\'e} \& Shield(2004)}]{aubeandshield}
\bibinfo{author}{Aub{\'e}, F.}, \& \bibinfo{author}{Shield, R.}
  (\bibinfo{year}{2004}).
\newblock \bibinfo{title}{Modeling the effect of leadership on crowd flow
  dynamics}.
\newblock In \bibinfo{editor}{P.~M.~A. Sloot}, \bibinfo{editor}{B.~Chopard}, \&
  \bibinfo{editor}{A.~G. Hoekstra} (Eds.), {\it \bibinfo{booktitle}{Cellular
  Automata}\/} (pp. \bibinfo{pages}{601--611}).
\newblock \bibinfo{publisher}{Springer}.
\newblock \DOIprefix\doi{https://doi.org/10.1007/978-3-540-30479-1_62}.
\bibitem[{Bertsekas \& Tsitsiklis(1996)}]{neurodynamic}
\bibinfo{author}{Bertsekas, D.~P.}, \& \bibinfo{author}{Tsitsiklis, J.~N.}
  (\bibinfo{year}{1996}).
\newblock {\it \bibinfo{title}{Neuro-dynamic programming}\/}
  volume~\bibinfo{volume}{5}.
\newblock \bibinfo{address}{Belmont, MA}: \bibinfo{publisher}{Athena
  Scientific}.
\bibitem[{Birge(1982)}]{birge1982}
\bibinfo{author}{Birge, J.~R.} (\bibinfo{year}{1982}).
\newblock \bibinfo{title}{The value of the stochastic solution in stochastic
  linear programs with fixed recourse}.
\newblock {\it \bibinfo{journal}{Mathematical Programming}\/},  {\it
  \bibinfo{volume}{24}\/}, \bibinfo{pages}{314--325}.
  \DOIprefix\doi{https://doi.org/10.1007/BF01585113}.
\bibitem[{Calafiore \& Campi(2006)}]{scenarioapproach}
\bibinfo{author}{Calafiore, G.~C.}, \& \bibinfo{author}{Campi, M.~C.}
  (\bibinfo{year}{2006}).
\newblock \bibinfo{title}{The scenario approach to robust control design}.
\newblock {\it \bibinfo{journal}{IEEE Transactions on Automatic Control}\/},
  {\it \bibinfo{volume}{51}\/}, \bibinfo{pages}{742--753}.
  \DOIprefix\doi{https://doi.org/10.1109/TAC.2006.875041}.
\bibitem[{Cao et~al.(2016)Cao, Song \& Lv}]{multigridmodel}
\bibinfo{author}{Cao, S.}, \bibinfo{author}{Song, W.}, \& \bibinfo{author}{Lv,
  W.} (\bibinfo{year}{2016}).
\newblock \bibinfo{title}{Modeling pedestrian evacuation with guiders based on
  a multi-grid model}.
\newblock {\it \bibinfo{journal}{Physics Letters A}\/},  {\it
  \bibinfo{volume}{380}\/}, \bibinfo{pages}{540--547}.
  \DOIprefix\doi{https://doi.org/10.1016/j.physleta.2015.11.028}.
\bibitem[{Cristiani \& Peri(2017)}]{obstacles}
\bibinfo{author}{Cristiani, E.}, \& \bibinfo{author}{Peri, D.}
  (\bibinfo{year}{2017}).
\newblock \bibinfo{title}{Handling obstacles in pedestrian simulations: Models
  and optimization}.
\newblock {\it \bibinfo{journal}{Applied Mathematical Modelling}\/},  {\it
  \bibinfo{volume}{45}\/}, \bibinfo{pages}{285--302}.
  \DOIprefix\doi{https://doi.org/10.1016/j.apm.2016.12.020}.
\bibitem[{Dyer et~al.(2009)Dyer, Johansson, Helbing, Couzin \&
  Krause}]{humangroup}
\bibinfo{author}{Dyer, J. R.~G.}, \bibinfo{author}{Johansson, A.},
  \bibinfo{author}{Helbing, D.}, \bibinfo{author}{Couzin, I.~D.}, \&
  \bibinfo{author}{Krause, J.} (\bibinfo{year}{2009}).
\newblock \bibinfo{title}{Leadership, consensus decision making and collective
  behaviour in humans}.
\newblock {\it \bibinfo{journal}{Philosophical Transactions of the Royal
  Society B: Biological Sciences}\/},  {\it \bibinfo{volume}{364}\/},
  \bibinfo{pages}{781--789}.
  \DOIprefix\doi{https://doi.org/10.1098/rstb.2008.0233}.
\bibitem[{Goldberg(1989)}]{goldberg}
\bibinfo{author}{Goldberg, D.~E.} (\bibinfo{year}{1989}).
\newblock {\it \bibinfo{title}{Genetic Algorithms in Search, Optimization and
  Machine Learning}\/}.
\newblock \bibinfo{address}{Boston, MA, USA}:
  \bibinfo{publisher}{Addison-Wesley Longman Publishing Co.}
\bibitem[{Gwynne et~al.(2016)Gwynne, Hulse \& Kinsey}]{gwynne2016}
\bibinfo{author}{Gwynne, S. M.~V.}, \bibinfo{author}{Hulse, L.~M.}, \&
  \bibinfo{author}{Kinsey, M.~J.} (\bibinfo{year}{2016}).
\newblock \bibinfo{title}{Guidance for the model developer on representing
  human behavior in egress models}.
\newblock {\it \bibinfo{journal}{Fire Technology}\/},  {\it
  \bibinfo{volume}{52}\/}, \bibinfo{pages}{775--800}.
  \DOIprefix\doi{https://doi.org/10.1007/s10694-015-0501-2}.
\bibitem[{Haghani(2020)}]{haghani2020optimising}
\bibinfo{author}{Haghani, M.} (\bibinfo{year}{2020}).
\newblock \bibinfo{title}{Optimising crowd evacuations: Mathematical,
  architectural and behavioural approaches}.
\newblock {\it \bibinfo{journal}{Safety Science}\/},  {\it
  \bibinfo{volume}{128}\/}, \bibinfo{pages}{104745}.
  \DOIprefix\doi{https://doi.org/10.1016/j.ssci.2020.104745}.
\bibitem[{Hammel \& B\"{a}ck(1994)}]{noisyfunction}
\bibinfo{author}{Hammel, U.}, \& \bibinfo{author}{B\"{a}ck, T.}
  (\bibinfo{year}{1994}).
\newblock \bibinfo{title}{Evolution strategies on noisy functions: How to
  improve convergence properties}.
\newblock In \bibinfo{editor}{Y.~Davidor}, \bibinfo{editor}{H.-P. Schwefel}, \&
  \bibinfo{editor}{R.~M{\"a}nner} (Eds.), {\it \bibinfo{booktitle}{Parallel
  Problem Solving from Nature--PPSN III}\/} (pp. \bibinfo{pages}{159--168}).
\newblock \bibinfo{publisher}{Springer}.
\newblock \DOIprefix\doi{https://doi.org/10.1007/3-540-58484-6_260}.
\bibitem[{Helbing et~al.(2003)Helbing, Isobe, Nagatani \&
  Takimoto}]{latticegassimulation}
\bibinfo{author}{Helbing, D.}, \bibinfo{author}{Isobe, M.},
  \bibinfo{author}{Nagatani, T.}, \& \bibinfo{author}{Takimoto, K.}
  (\bibinfo{year}{2003}).
\newblock \bibinfo{title}{Lattice gas simulation of experimentally studied
  evacuation dynamics}.
\newblock {\it \bibinfo{journal}{Physical Review E}\/},  {\it
  \bibinfo{volume}{67}\/}, \bibinfo{pages}{067101}.
  \DOIprefix\doi{https://doi.org/10.1103/PhysRevE.67.067101}.
\bibitem[{Helbing \& Johansson(2013)}]{helbing2013}
\bibinfo{author}{Helbing, D.}, \& \bibinfo{author}{Johansson, A.}
  (\bibinfo{year}{2013}).
\newblock \bibinfo{title}{Pedestrian, crowd, and evacuation dynamics}.
\newblock {\it \bibinfo{journal}{arXiv preprint}\/}, .
\newblock \bibinfo{note}{ArXiv:1309.1609}.
\bibitem[{Helbing \& Moln\'{a}r(1995)}]{socialforce95}
\bibinfo{author}{Helbing, D.}, \& \bibinfo{author}{Moln\'{a}r, P.}
  (\bibinfo{year}{1995}).
\newblock \bibinfo{title}{Social force model for pedestrian dynamics}.
\newblock {\it \bibinfo{journal}{Physical Review E}\/},  {\it
  \bibinfo{volume}{51}\/}, \bibinfo{pages}{4282}.
  \DOIprefix\doi{https://doi.org/10.1103/PhysRevE.51.4282}.
\bibitem[{Heli\"{o}vaara et~al.(2013)Heli\"{o}vaara, Ehtamo, Helbing \&
  Korhonen}]{ownphysreve}
\bibinfo{author}{Heli\"{o}vaara, S.}, \bibinfo{author}{Ehtamo, H.},
  \bibinfo{author}{Helbing, D.}, \& \bibinfo{author}{Korhonen, T.}
  (\bibinfo{year}{2013}).
\newblock \bibinfo{title}{Patient and impatient pedestrians in a spatial game
  for egress congestion}.
\newblock {\it \bibinfo{journal}{Physical Review E}\/},  {\it
  \bibinfo{volume}{87}\/}, \bibinfo{pages}{012802}.
  \DOIprefix\doi{https://doi.org/10.1103/PhysRevE.87.012802}.
\bibitem[{Heli{\"o}vaara et~al.(2012a)Heli{\"o}vaara, Korhonen, Hostikka \&
  Ehtamo}]{heliovaara2012counterflow}
\bibinfo{author}{Heli{\"o}vaara, S.}, \bibinfo{author}{Korhonen, T.},
  \bibinfo{author}{Hostikka, S.}, \& \bibinfo{author}{Ehtamo, H.}
  (\bibinfo{year}{2012a}).
\newblock \bibinfo{title}{Counterflow model for agent-based simulation of crowd
  dynamics}.
\newblock {\it \bibinfo{journal}{Building and Environment}\/},  {\it
  \bibinfo{volume}{48}\/}, \bibinfo{pages}{89--100}.
  \DOIprefix\doi{https://doi.org/10.1016/j.buildenv.2011.08.020}.
\bibitem[{Heli\"{o}vaara et~al.(2012b)Heli\"{o}vaara, Kuusinen, Rinne, Korhonen
  \& Ehtamo}]{safetyscience}
\bibinfo{author}{Heli\"{o}vaara, S.}, \bibinfo{author}{Kuusinen, J.-M.},
  \bibinfo{author}{Rinne, T.}, \bibinfo{author}{Korhonen, T.}, \&
  \bibinfo{author}{Ehtamo, H.} (\bibinfo{year}{2012b}).
\newblock \bibinfo{title}{Pedestrian behavior and exit selection in evacuation
  of a corridor--{A}n experimental study}.
\newblock {\it \bibinfo{journal}{Safety Science}\/},  {\it
  \bibinfo{volume}{50}\/}, \bibinfo{pages}{221--227}.
  \DOIprefix\doi{https://doi.org/10.1016/j.ssci.2011.08.020}.
\bibitem[{Hoogendoorn \& Bovy(2003)}]{differentialgames}
\bibinfo{author}{Hoogendoorn, S.}, \& \bibinfo{author}{Bovy, P. H.~L.}
  (\bibinfo{year}{2003}).
\newblock \bibinfo{title}{Simulation of pedestrian flows by optimal control and
  differential games}.
\newblock {\it \bibinfo{journal}{Optimal Control Applications and Methods}\/},
  {\it \bibinfo{volume}{24}\/}, \bibinfo{pages}{153--172}.
  \DOIprefix\doi{https://doi.org/10.1002/oca.727}.
\bibitem[{Hou et~al.(2014)Hou, Liu, Pan \& Wang}]{leadershipeffect}
\bibinfo{author}{Hou, L.}, \bibinfo{author}{Liu, J.-G.}, \bibinfo{author}{Pan,
  X.}, \& \bibinfo{author}{Wang, B.-H.} (\bibinfo{year}{2014}).
\newblock \bibinfo{title}{A social force evacuation model with the leadership
  effect}.
\newblock {\it \bibinfo{journal}{Physica A: Statistical Mechanics and its
  Applications}\/},  {\it \bibinfo{volume}{400}\/}, \bibinfo{pages}{93--99}.
  \DOIprefix\doi{https://doi.org/10.1016/j.physa.2013.12.049}.
\bibitem[{Huang \& Guo(2008)}]{staticfloorfield}
\bibinfo{author}{Huang, H.-J.}, \& \bibinfo{author}{Guo, R.-Y.}
  (\bibinfo{year}{2008}).
\newblock \bibinfo{title}{Static floor field and exit choice for pedestrian
  evacuation in rooms with internal obstacles and multiple exits}.
\newblock {\it \bibinfo{journal}{Physical Review E}\/},  {\it
  \bibinfo{volume}{78}\/}, \bibinfo{pages}{021131}.
  \DOIprefix\doi{https://doi.org/10.1103/PhysRevE.78.021131}.
\bibitem[{Karamouzas et~al.(2014)Karamouzas, Skinner \& Guy}]{powerlaw}
\bibinfo{author}{Karamouzas, I.}, \bibinfo{author}{Skinner, B.}, \&
  \bibinfo{author}{Guy, S.~J.} (\bibinfo{year}{2014}).
\newblock \bibinfo{title}{Universal power law governing pedestrian
  interactions}.
\newblock {\it \bibinfo{journal}{Physical Review Letters}\/},  {\it
  \bibinfo{volume}{113}\/}, \bibinfo{pages}{238701}.
  \DOIprefix\doi{https://doi.org/10.1103/PhysRevLett.113.238701}.
\bibitem[{Karamouzas et~al.(2017)Karamouzas, Sohre, Narain \&
  Guy}]{implicitcrowd}
\bibinfo{author}{Karamouzas, I.}, \bibinfo{author}{Sohre, N.},
  \bibinfo{author}{Narain, R.}, \& \bibinfo{author}{Guy, S.~J.}
  (\bibinfo{year}{2017}).
\newblock \bibinfo{title}{Implicit crowds: Optimization integrator for robust
  crowd simulation}.
\newblock {\it \bibinfo{journal}{ACM Transactions on Graphics}\/},  {\it
  \bibinfo{volume}{36}\/}, \bibinfo{pages}{1--13}.
  \DOIprefix\doi{https://doi.org/10.1145/3072959.3073705}.
\bibitem[{Korhonen(2018)}]{fdsevac}
\bibinfo{author}{Korhonen, T.} (\bibinfo{year}{2018}).
\newblock {\it \bibinfo{title}{Fire dynamics simulator with evacuation:
  {FDS}+{E}vac. Technical Reference and User's Guide}\/}.
\newblock \bibinfo{type}{Technical Report} VTT Technical Research Centre of
  Finland.
\newblock \URLprefix
  \url{http://virtual.vtt.fi/virtual/proj6/fdsevac/documents/FDS+EVAC_Guide.pdf}.
\bibitem[{Kretz et~al.(2011)Kretz, Grosse, Hengst, Kautzsch, Pohlmann \&
  Vortisch}]{kretz}
\bibinfo{author}{Kretz, T.}, \bibinfo{author}{Grosse, A.},
  \bibinfo{author}{Hengst, S.}, \bibinfo{author}{Kautzsch, L.},
  \bibinfo{author}{Pohlmann, A.}, \& \bibinfo{author}{Vortisch, P.}
  (\bibinfo{year}{2011}).
\newblock \bibinfo{title}{Quickest paths in simulations of pedestrians}.
\newblock {\it \bibinfo{journal}{Advances in Complex Systems}\/},  {\it
  \bibinfo{volume}{14}\/}, \bibinfo{pages}{733--759}.
  \DOIprefix\doi{https://doi.org/10.1142/S0219525911003281}.
\bibitem[{Kurdi et~al.(2018)Kurdi, Al-Megren, Althunyan \&
  Almulifi}]{kurdi2018effect}
\bibinfo{author}{Kurdi, H.~A.}, \bibinfo{author}{Al-Megren, S.},
  \bibinfo{author}{Althunyan, R.}, \& \bibinfo{author}{Almulifi, A.}
  (\bibinfo{year}{2018}).
\newblock \bibinfo{title}{Effect of exit placement on evacuation plans}.
\newblock {\it \bibinfo{journal}{European Journal of Operational Research}\/},
  {\it \bibinfo{volume}{269}\/}, \bibinfo{pages}{749--759}.
  \DOIprefix\doi{https://doi.org/10.1016/j.ejor.2018.01.050}.
\bibitem[{Li et~al.(2016)Li, Li, Gong \& Shen}]{tracemodel}
\bibinfo{author}{Li, W.}, \bibinfo{author}{Li, Y.}, \bibinfo{author}{Gong, J.},
  \& \bibinfo{author}{Shen, S.} (\bibinfo{year}{2016}).
\newblock \bibinfo{title}{The {T}race {M}odel: A model for simulation of the
  tracing process during evacuations in complex route environments}.
\newblock {\it \bibinfo{journal}{Simulation Modelling Practice and Theory}\/},
  {\it \bibinfo{volume}{60}\/}, \bibinfo{pages}{108--121}.
  \DOIprefix\doi{https://doi.org/10.1016/j.simpat.2015.09.011}.
\bibitem[{L{\o}v{\aa}s(1995)}]{lovos}
\bibinfo{author}{L{\o}v{\aa}s, G.} (\bibinfo{year}{1995}).
\newblock \bibinfo{title}{On performance measures for evacuation systems}.
\newblock {\it \bibinfo{journal}{European Journal of Operational Research}\/},
  {\it \bibinfo{volume}{85}\/}, \bibinfo{pages}{352--367}.
  \DOIprefix\doi{https://doi.org/10.1016/0377-2217(94)00054-G}.
\bibitem[{L{\o}v{\aa}s(1998)}]{lovosflow}
\bibinfo{author}{L{\o}v{\aa}s, G.} (\bibinfo{year}{1998}).
\newblock \bibinfo{title}{Models of wayfinding in emergency evacuations}.
\newblock {\it \bibinfo{journal}{European Journal of Operational Research}\/},
  {\it \bibinfo{volume}{105}\/}, \bibinfo{pages}{371--389}.
  \DOIprefix\doi{https://doi.org/10.1016/S0377-2217(97)00084-2}.
\bibitem[{McCormack \& Chen(2014)}]{optimizingproportion}
\bibinfo{author}{McCormack, P.}, \& \bibinfo{author}{Chen, T.-Y.}
  (\bibinfo{year}{2014}).
\newblock \bibinfo{title}{Optimizing leader proportion and behavior for
  evacuating buildings}.
\newblock In {\it \bibinfo{booktitle}{Proceedings of the 2014 Symposium on
  Agent Directed Simulation}\/} ADS '14 (pp. \bibinfo{pages}{13:1--13:6}).
\newblock \bibinfo{publisher}{Society for Computer Simulation International}.
\bibitem[{Nishinari et~al.(2004)Nishinari, Kirchner, Namazi \&
  Schadschneider}]{nishinari}
\bibinfo{author}{Nishinari, K.}, \bibinfo{author}{Kirchner, A.},
  \bibinfo{author}{Namazi, A.}, \& \bibinfo{author}{Schadschneider, A.}
  (\bibinfo{year}{2004}).
\newblock \bibinfo{title}{Extended floor field {CA} model for evacuation
  dynamics}.
\newblock {\it \bibinfo{journal}{IEICE Transactions on Information and
  Systems}\/},  {\it \bibinfo{volume}{87}\/}, \bibinfo{pages}{726--732}.
\bibitem[{NIST(2020)}]{nist}
\bibinfo{author}{NIST} (\bibinfo{year}{2020}).
\newblock \bibinfo{title}{Fire dynamics simulator}.
\newblock \URLprefix \url{https://pages.nist.gov/fds-smv/}
  \bibinfo{note}{accessed: 2020-03-16}.
\bibitem[{Oliphant et~al.(2020)}]{numba}
\bibinfo{author}{Oliphant, T.} et~al. (\bibinfo{year}{2020}).
\newblock \bibinfo{title}{{Numba}: A high performance python compiler}.
\newblock \URLprefix \url{https://numba.pydata.org} \bibinfo{note}{accessed:
  2020-03-16}.
\bibitem[{Pelechano \& Badler(2006)}]{trainedleader}
\bibinfo{author}{Pelechano, N.}, \& \bibinfo{author}{Badler, N.~I.}
  (\bibinfo{year}{2006}).
\newblock \bibinfo{title}{Modeling crowd and trained leader behavior during
  building evacuation}.
\newblock {\it \bibinfo{journal}{IEEE Computer Graphics and Applications}\/},
  {\it \bibinfo{volume}{26}\/}, \bibinfo{pages}{80--86}.
  \DOIprefix\doi{https://doi.org/10.1109/MCG.2006.133}.
\bibitem[{Proulx(2002)}]{proulx2002}
\bibinfo{author}{Proulx, G.} (\bibinfo{year}{2002}).
\newblock \bibinfo{title}{Movement of people: the evacuation timing}.
\newblock In \bibinfo{editor}{P.~J. DiNenno} et~al. (Eds.), {\it
  \bibinfo{booktitle}{SFPE Handbook of Fire Protection Engineering}\/} (pp.
  \bibinfo{pages}{342--366}).
\newblock \bibinfo{publisher}{the National Fire Protection Association}.
\bibitem[{Saadatseresht et~al.(2009)Saadatseresht, Mansourian \& Taleai}]{moea}
\bibinfo{author}{Saadatseresht, M.}, \bibinfo{author}{Mansourian, A.}, \&
  \bibinfo{author}{Taleai, M.} (\bibinfo{year}{2009}).
\newblock \bibinfo{title}{Evacuation planning using multiobjective evolutionary
  optimization approach}.
\newblock {\it \bibinfo{journal}{European Journal of Operational Research}\/},
  {\it \bibinfo{volume}{198}\/}, \bibinfo{pages}{305--314}.
  \DOIprefix\doi{https://doi.org/10.1016/j.ejor.2008.07.032}.
\bibitem[{Schadschneider et~al.(2008)Schadschneider, Klingsch, Kl{\"u}pfel,
  Kretz, Rogsch \& Seyfried}]{schadschneider}
\bibinfo{author}{Schadschneider, A.}, \bibinfo{author}{Klingsch, W.},
  \bibinfo{author}{Kl{\"u}pfel, H.}, \bibinfo{author}{Kretz, T.},
  \bibinfo{author}{Rogsch, C.}, \& \bibinfo{author}{Seyfried, A.}
  (\bibinfo{year}{2008}).
\newblock \bibinfo{title}{Evacuation dynamics: Empirical results, modeling and
  applications}.
\newblock {\it \bibinfo{journal}{arXiv preprint}\/}, .
\newblock \bibinfo{note}{ArXiv:0802.1620}.
\bibitem[{von Schantz(2020)}]{articlecode}
\bibinfo{author}{von Schantz, A.} (\bibinfo{year}{2020}).
\newblock \bibinfo{title}{{Minimizing the evacuation time of a crowd from a
  complex building using rescue guides – code (Version 1.2)}}.
\newblock \bibinfo{howpublished}{Zenodo}.
\newblock \DOIprefix\doi{10.5281/zenodo.3831338}.
\bibitem[{von Schantz \& Ehtamo(2015)}]{ownca}
\bibinfo{author}{von Schantz, A.}, \& \bibinfo{author}{Ehtamo, H.}
  (\bibinfo{year}{2015}).
\newblock \bibinfo{title}{Spatial game in cellular automaton evacuation model}.
\newblock {\it \bibinfo{journal}{Physical Review E}\/},  {\it
  \bibinfo{volume}{92}\/}, \bibinfo{pages}{052805}.
  \DOIprefix\doi{https://doi.org/10.1103/PhysRevE.92.052805}.
\bibitem[{von Schantz \& Ehtamo(2019)}]{ownphysica}
\bibinfo{author}{von Schantz, A.}, \& \bibinfo{author}{Ehtamo, H.}
  (\bibinfo{year}{2019}).
\newblock \bibinfo{title}{Pushing and overtaking others in a spatial game of
  exit congestion}.
\newblock {\it \bibinfo{journal}{Physica A: Statistical Mechanics and its
  Applications}\/},  {\it \bibinfo{volume}{527}\/}, \bibinfo{pages}{121151}.
  \DOIprefix\doi{https://doi.org/10.1016/j.physa.2019.121151}.
\bibitem[{Sethian(1999)}]{fastmarchingmethod}
\bibinfo{author}{Sethian, J.~A.} (\bibinfo{year}{1999}).
\newblock {\it \bibinfo{title}{Level set methods and fast marching methods:
  Evolving interfaces in computational geometry, fluid mechanics, computer
  vision, and materials science}\/} volume~\bibinfo{volume}{3}.
\newblock \bibinfo{publisher}{Cambridge University Press}.
\bibitem[{Vermuyten et~al.(2016)Vermuyten, Beli\"{e}n, De~Boeck, Reniers \&
  Wauters}]{vermuyten}
\bibinfo{author}{Vermuyten, H.}, \bibinfo{author}{Beli\"{e}n, J.},
  \bibinfo{author}{De~Boeck, L.}, \bibinfo{author}{Reniers, G.}, \&
  \bibinfo{author}{Wauters, T.} (\bibinfo{year}{2016}).
\newblock \bibinfo{title}{A review of optimisation models for pedestrian
  evacuation and design problems}.
\newblock {\it \bibinfo{journal}{Safety Science}\/},  {\it
  \bibinfo{volume}{87}\/}, \bibinfo{pages}{167--178}.
  \DOIprefix\doi{https://doi.org/10.1016/j.ssci.2016.04.001}.
\bibitem[{Wang et~al.(2012)Wang, Zheng \& Cheng}]{evacuationassistants}
\bibinfo{author}{Wang, X.}, \bibinfo{author}{Zheng, X.}, \&
  \bibinfo{author}{Cheng, Y.} (\bibinfo{year}{2012}).
\newblock \bibinfo{title}{Evacuation assistants: An extended model for
  determining effective locations and optimal numbers}.
\newblock {\it \bibinfo{journal}{Physica A: Statistical Mechanics and its
  Applications}\/},  {\it \bibinfo{volume}{391}\/},
  \bibinfo{pages}{2245--2260}.
  \DOIprefix\doi{https://doi.org/10.1016/j.physa.2011.11.051}.
\bibitem[{Weiss et~al.(2019)Weiss, Litteneker, Jiang \&
  Terzopoulos}]{positionbased}
\bibinfo{author}{Weiss, T.}, \bibinfo{author}{Litteneker, A.},
  \bibinfo{author}{Jiang, C.}, \& \bibinfo{author}{Terzopoulos, D.}
  (\bibinfo{year}{2019}).
\newblock \bibinfo{title}{Position-based real-time simulation of large crowds}.
\newblock {\it \bibinfo{journal}{Computers \& Graphics}\/},  {\it
  \bibinfo{volume}{78}\/}, \bibinfo{pages}{12--22}.
  \DOIprefix\doi{https://doi.org/10.1016/j.cag.2018.10.008}.
\bibitem[{Yao et~al.(2004)Yao, Wang, Liu \& Cheng}]{blocklist}
\bibinfo{author}{Yao, Z.}, \bibinfo{author}{Wang, J.~S.}, \bibinfo{author}{Liu,
  G.~R.}, \& \bibinfo{author}{Cheng, M.} (\bibinfo{year}{2004}).
\newblock \bibinfo{title}{Improved neighbor list algorithm in molecular
  simulations using cell decomposition and data sorting method}.
\newblock {\it \bibinfo{journal}{Computer Physics Communications}\/},  {\it
  \bibinfo{volume}{161}\/}, \bibinfo{pages}{27--35}.
  \DOIprefix\doi{https://doi.org/10.1016/j.cpc.2004.04.004}.
\bibitem[{Zhou et~al.(2019)Zhou, Dong, Zhao, Ioannou \&
  Wang}]{maximumcoverage2}
\bibinfo{author}{Zhou, M.}, \bibinfo{author}{Dong, H.}, \bibinfo{author}{Zhao,
  Y.}, \bibinfo{author}{Ioannou, P.~A.}, \& \bibinfo{author}{Wang, F.-Y.}
  (\bibinfo{year}{2019}).
\newblock \bibinfo{title}{Optimization of crowd evacuation with leaders in
  urban rail transit stations}.
\newblock {\it \bibinfo{journal}{IEEE Transactions on Intelligent
  Transportation Systems}\/},  {\it \bibinfo{volume}{20}\/},
  \bibinfo{pages}{4476--4487}.
  \DOIprefix\doi{https://doi.org/10.1109/TITS.2018.2886415}.

\end{thebibliography}

\end{document}